\documentclass[twocolumn]{aastex631}

\newcommand\tnote[1]{$^{\rm #1}$}

\newcommand{\kam}[1]{\textcolor{red}{[KAM]}}

\usepackage{amsmath}
\usepackage{multirow}
\usepackage{threeparttable}
\usepackage{ulem}
\usepackage{soul}
\usepackage{longtable}
\usepackage{comment}

\usepackage{tabularx} 
\usepackage{booktabs} 

\begin{document}

\title{Monitoring of 3C~286 with ALMA, IRAM, and SMA from 2006 to 2025: Stability, Synchrotron Ages, and Frequency-Dependent Polarization Attributed to Core-Shift}

\correspondingauthor{Minchul Kam}
\email{mkam@asiaa.sinica.edu.tw}

\author[0000-0001-9799-765X]{Minchul Kam}
\affiliation{\normalfont Institute of Astronomy and Astrophysics, Academia Sinica, P.O. Box 23-141, Taipei 10617, Taiwan}
\affiliation{\normalfont Department of Physics and Astronomy, Seoul National University, Gwanak-gu, Seoul 08826, Republic of Korea}

\author[0000-0003-0292-3645]{Hiroshi Nagai}
\affiliation{\normalfont National Astronomical Observatory of Japan, 2-21-1 Osawa, Mitaka, Tokyo 181-8588, Japan}
\affiliation{\normalfont Department of Astronomical Science, The Graduate University for Advanced Studies, SOKENDAI, 2-21-1 Osawa, Mitaka, Tokyo 181-8588, Japan}

\author[0000-0002-2709-7338]{Motoki Kino}
\affiliation{\normalfont National Astronomical Observatory of Japan, 2-21-1 Osawa, Mitaka, Tokyo 181-8588, Japan}
\affiliation{\normalfont Kogakuin University of Technology $\&$ Engineering, Academic Support Center, 2665-1 Nakano, Hachioji, Tokyo 192-0015, Japan}

\author[0000-0001-6988-8763]{Keiichi Asada}
\affiliation{\normalfont Institute of Astronomy and Astrophysics, Academia Sinica, P.O. Box 23-141, Taipei 10617, Taiwan}

\author[0000-0002-5580-006X]{R{\"u}diger Kneissl}
\affiliation{\normalfont Joint ALMA Observatory, Alonso de Cordova 3107, Vitacura, Santiago 763-0355, Chile}
\affiliation{\normalfont European Southern Observatory, Alonso de Córdova 3107, Casilla 19, Santiago, Chile}

\author[0000-0002-3777-6182]{Iv\'an Agudo}
\affiliation{\normalfont Instituto de Astrof\'isica de Andaluc\'ia-CSIC, Glorieta de la Astronom\'ia, E-18008, Granada, Spain}

\author[0000-0003-0465-1559]{Sascha Trippe}
\affiliation{\normalfont Department of Physics and Astronomy, Seoul National University, Gwanak-gu, Seoul 08826, Republic of Korea}
\affiliation{\normalfont SNU Astronomy Research Center, Seoul National University, Gwanak-gu, Seoul 08826, Republic of Korea
}

\author[0000-0002-5158-0063]{Seiji Kameno}
\affiliation{\normalfont National Astronomical Observatory of Japan, 2-21-1 Osawa, Mitaka, Tokyo 181-8588, Japan}
\affiliation{\normalfont Joint ALMA Observatory, Alonso de Cordova 3107, Vitacura, Santiago 763-0355, Chile}

\author[0000-0003-3025-9497]{Ioannis Myserlis}
\affiliation{\normalfont Institut de Radiostronomie Milim\'etrique, Avenida Divina Pastora, 7, Local 20, E-18012 Granada, Spain}

\author[0000-0002-1407-7944]{Ramprasad Rao}
\affiliation{\normalfont Center for Astrophysics $|$ Harvard \& Smithsonian, 60 Garden Street, Cambridge, MA 02138, USA}

\author[0000-0003-2010-8521]{Hojin Cho}
\affiliation{\normalfont Department of Physics \& Astronomy, Texas Tech University, Lubbock TX, 79409-1051, USA}

\author[0000-0001-7097-8360]{Richard A. Perley}
\affiliation{\normalfont National Radio Astronomy Observatory, 1003 Lopezville Road, Socorro, NM 87801, USA}

\author[0000-0002-5344-820X]{Bryan J. Butler}
\affiliation{\normalfont National Radio Astronomy Observatory, 1003 Lopezville Road, Socorro, NM 87801, USA}

\author[0000-0003-0685-3621]{Mark Gurwell}
\affiliation{\normalfont Center for Astrophysics $|$ Harvard \& Smithsonian, 60 Garden Street, Cambridge, MA 02138, USA}

\author[0000-0002-6916-3559]{Tomoki Matsuoka}
\affiliation{\normalfont Institute of Astronomy and Astrophysics, Academia Sinica, P.O. Box 23-141, Taipei 10617, Taiwan}

\author[0000-0001-6558-9053]{Jongho Park}
\affiliation{\normalfont School of Space Research, Kyung Hee University, 1732, Deogyeong-daero, Giheung-gu, Yongin-si, Gyeonggi-do 17104, Republic of Korea}
\affiliation{\normalfont Institute of Astronomy and Astrophysics, Academia Sinica, P.O. Box 23-141, Taipei 10617, Taiwan}

\author[0000-0003-1117-2863]{Carolina Casadio}
\affiliation{\normalfont Institute of Astrophysics, Foundation for Research and Technology-Hellas, Voutes, 7110 Heraklion, Greece}
\affiliation{\normalfont Department of Physics, University of Crete, 70013, Heraklion, Greece}

\author{Baltasar Vila Vilaro}
\affiliation{\normalfont Joint ALMA Observatory, Alonso de Cordova 3107, Vitacura, Santiago 763-0355, Chile}
\affiliation{\normalfont European Southern Observatory, Alonso de Córdova 3107, Casilla 19, Santiago, Chile}

\author{Celia Verdugo}
\affiliation{\normalfont Joint ALMA Observatory, Alonso de Cordova 3107, Vitacura, Santiago 763-0355, Chile}

\author{Matias Radiszcz}
\affiliation{\normalfont Joint ALMA Observatory, Alonso de Cordova 3107, Vitacura, Santiago 763-0355, Chile}

\author{Kurt Plarre}
\affiliation{\normalfont Joint ALMA Observatory, Alonso de Cordova 3107, Vitacura, Santiago 763-0355, Chile}

\author[0009-0008-9101-0330]{Wanchaloem Khwammai}
\affiliation{\normalfont Department of Physics and Astronomy, Seoul National University, Gwanak-gu, Seoul 08826, Republic of Korea}

\author{Diego \'Alvarez-Ortega}
\affiliation{\normalfont Institute of Astrophysics, Foundation for Research and Technology-Hellas, Voutes, 7110 Heraklion, Greece}
\affiliation{\normalfont Department of Physics, University of Crete, 70013, Heraklion, Greece}

\author[0000-0002-4131-655X]{Juan Escudero}
\affiliation{\normalfont Center for Astrophysics $|$ Harvard \& Smithsonian, 60 Garden Street, Cambridge, MA 02138, USA}

\author[0009-0006-5292-6974]{Clemens Thum}
\affiliation{\normalfont Institut de Radiostronomie Milim\'etrique, Avenida Divina Pastora, 7, Local 20, E-18012 Granada, Spain}

\author[0000-0002-3490-146X]{Garrett Keating}
\affiliation{\normalfont Center for Astrophysics $|$ Harvard \& Smithsonian, 60 Garden Street, Cambridge, MA 02138, USA}

\begin{abstract}
\noindent We present the results of multi-frequency monitoring of the radio quasar 3C~286, conducted using three instruments: ALMA at 91.5, 103.5, 233.0, and 343.4~GHz, the IRAM 30-m Telescope at 86 and 229~GHz, and SMA at 225~GHz. 
The IRAM measurements from 2006 to 2024 show that the total flux of 3C~286 is stable within measurement uncertainties, indicating long-term stability up to 229~GHz, when applying a fixed Kelvin-to-Jansky conversion factor throughout its dataset. ALMA data from 2018 to 2024 exhibit a decrease in flux, which up to 4\% could be attributed to an apparent increase in the absolute brightness of Uranus, the primary flux calibrator for ALMA with the ESA4 model. Taken together, these results suggest that the intrinsic total flux of 3C~286 has remained stable up to 229~GHz over the monitoring period. 
The polarization properties of 3C~286 are stable across all observing frequencies. The electric vector position angle (EVPA) gradually rotates as a function of wavelength squared, which is well described by a single power-law over the full frequency range. We therefore propose using the theoretical EVPA values from this model curve for absolute EVPA calibration between 5 and 343.4 GHz. The Faraday rotation measure increases as a function of frequency up to $\rm (3.2\pm1.5)\times10^4\ rad\ m^{-2}$, following $\mid$RM$\mid\propto \nu$\textsuperscript{$\alpha$} with $\alpha=2.05\pm0.06$. This trend is consistent with the core-shift effect expected in a conical jet.
\end{abstract}


\section{Introduction} \label{sec:intro}

\noindent 3C~286 is a quasar ($z=0.849$; \citet{burbidge69}) classified as a compact steep-spectrum (CSS) source\footnote{Recent studies suggest that 3C~286 can also be classified as a narrow-line Seyfert 1 (NLS1) galaxy \citep[e.g.,][]{yao21}.} \citep{peacock82, fanti85, odea98}, characterized by a compact size of less than 15~kpc and spectra that peak at frequencies typically below 400~MHz \citep[see][for reviews]{odea98, odea21}. It has been one of the most widely used calibrator sources in radio observations, as it best satisfies the three key criteria for an ideal calibrator: brightness, compactness, and stability. Specifically, it shows a compact structure at arcsecond scales \citep[e.g.,][]{kus88, spencer89, akujor91, akujor95, an04}, while exhibiting extended jets in the southwest direction when resolved with Very Long Baseline Interferometry (VLBI) observations at milliarcsecond scales \citep[e.g.,][]{wilkinson79, simon80, zhang94, jiang96, cotton97, hirabayashi00, fomalont00, an17}. 

The flux density of 3C~286 at 0.3--50~GHz has remained stable within a few percent over the past 50~years \citep{baars77, ott94, scaife12, perley13_flux, perley17, myserlis18, angelakis19, komossa24}. Monitoring observations of 3C~286 with the Very Large Array (VLA) from the 1980s to the 2010s show that the total flux reaches $\sim$30~Jy at $\sim$0.1~GHz and decrease at higher frequencies, while remaining stable over time within 1\% \citep{perley13_flux, perley17}. Notably, recent Effelsberg observations of 3C~286 at 2.6--41.3~GHz in 2023 November \citep{angelakis19, komossa24} confirmed that their total flux measurements at centimeter wavelengths are consistent not only with VLA measurements from the 1980s to the 2010s \citep{perley13_flux, perley17} but also with other measurements at 1.4--22.2~GHz from the 1970s \citep{baars77}, all within the Effelsberg measurement uncertainties. 
This suggests that the total flux of 3C~286 has been stable at these frequencies for the past 50 years. 

Similarly, the fractional polarization has also been stable at 1-50~GHz for a similar period. Polarimetric monitoring of 3C~286 using the VLA from 1980s to 2010s found that the fractional polarization remained stable within $2\%$ \citep{perley13_pol}. Specifically, it is 9.5\% at 1.5~GHz and gradually increases with frequency, reaching 13.1\% at 43.3~GHz \citep{perley13_pol}. Effelsberg observations also confirmed the increase in fractional polarization as a function of frequency, from 10.9\% at 2.6~GHz to 12.9\% at 14.3~GHz \citep{myserlis18, komossa24}. These values are consistent with the VLA measurements within a few percent, further supporting the stability of the fractional polarization of 3C~286. 

The electric vector position angle (EVPA)\footnote{In this paper, the EVPA is defined to increase from north toward east, following the IAU convention} \citep{iau74}. is also known to be stable for a similar period and frequency range \citep{perley13_pol, komossa24}. In addition, it appears to gradually rotate as frequency increases. For example, polarimetric observations using the 46-meter Algonquin Radio Telescope measured $\chi=33.0^\circ\pm0.9^\circ$ at 6.7 GHz and $\chi=31.0^\circ\pm1.3^\circ$ at 10.7 GHz \citep{bignell73}. Using the Very Large Array (VLA), \citet{perley13_pol} measured the EVPA of 3C~286 at frequencies from 5 to 45~GHz. Specifically, the EVPA values at five representative frequencies, averaged at each frequency from 1995 to 2012, gradually rotates as a function of frequency, from $33^\circ$ at 5~GHz to $35.8^\circ\pm0.1^\circ$ at 45 GHz \citep[see Table~2 in][]{perley13_pol}. This trend is consistent with measurements at both lower and higher frequencies, such as $29.2^\circ\pm0.1^\circ$ at 1.4~GHz with the MeerKAT telescope \citep{taylor24} and $37.3^\circ\pm0.8^\circ$ at 86 GHz with the IRAM Pico Veleta 30-meter Telescope (IRAM) \citep{agudo12}. 

However, the stability of 3C~286 at higher frequencies is less well understood, primarily because the brightness becomes significantly fainter. For example, at 229~GHz, the total flux drops to 0.3~Jy and the polarized flux to 0.04~Jy \citep[e.g.,][]{agudo12}, leading to larger measurement uncertainties. As a result, the EVPA values at this frequency measured using instruments such as the IRAM 30-meter Telescope \citep{agudo12}, the Submillimeter Array (SMA) \citep{marrone06, hull16}, the Combined Array for Research in Millimeter-wave Astronomy (CARMA) \citep{hull14, hull15}, and the Atacama Large Millimeter/submillimeter Array (ALMA) \citep{nagai16, kam23} are not well constrained and are distributed over a relatively wide range, from $33^\circ$ to 41$^\circ$. Similarly, the fractional polarization values measured with these instruments also show considerable scatter, ranging from 11\% to 17\%. These substantial uncertainties in EVPA and fractional polarization measurements at such high frequencies have made it challenging to investigate their stability and frequency-dependent trends \citep[see Figure 3 in][]{hull16}. 

In this paper, we present the result of polarimetric observations of 3C~286 using ALMA, IRAM, and SMA at 91.5--343.4~GHz, spanning from October 2006 to February 2025. The combined data from these three monitoring programs allows for a precise investigation of the stability and properties of 3C~286 at millimeter and submillimeter wavelengths. In Section~\ref{sec:Data}, we summarize the data used for the analysis. Results are shown in Section~\ref{sec:analysis}, and we discuss them in Section~\ref{sec:Discussions}. A Summary and conclusions follow in Section~\ref{sec:Conclusions}. We adopt cosmological parameters $H_{\rm 0}= \rm 73.30\ km\ s^{-1}\ Mpc^{-1}$ and $q_{0}=-0.51\pm0.024$ obtained from the Type Ia supernovae in the local universe \citep{riess22}. These parameters correspond to $\Omega_{\rm m}=\rm 0.327$ and $\Omega_{\rm \Lambda}=\rm 0.673$ in a flat universe. At a redshift of $z$=0.849, this leads to a linear scale of 1 mas = 7.23~pc.

\section{Data} \label{sec:Data}
\subsection{ALMA} \label{subsec:Data_alma}
\noindent 

About 40 bright calibrators, known as ``grid sources" and evenly distributed across the sky, are observed with the 7-m Array at least every two weeks to provide secondary flux calibrators in addition to solar system objects \citep{kneissl23}. 3C~286, a point-source calibrator for the ALMA 7-m Array, has been monitored as one of the grid sources\footnote{However, 3C 286 is at low elevations for ALMA (with the maximum elevation of 37$^\circ$) and its flux is relatively weak at higher frequencies. In addition, its flux calibration primarily relies on flux transfer from quasars, as its right ascension is opposite to the main planetary primary calibrators. As a result, it is more difficult to monitor than typical grid sources and is not used as a flux calibrator for ALMA, except in Band~1, where limited calibrator options make it a primary calibrator by necessity.}. The observations of 3C~286 were performed with a full polarization mode to obtain all four Stokes parameters. The data has been primarily obtained in Band~3 (91.5 and 103.5~GHz) and Band~7 (343.4~GHz) since May 2018. Data in Band~6 (233.0~GHz) has been obtained less frequently as a complement to the Band~7 data \citep{remijan19}. The observing frequency and monitoring periods are summarized in Table~\ref{tab:obs}. 

Observations of the grid sources include a scan of any available solar system object (SSO), with Uranus serving as the primary reference \citep{remijan19}. The CASA brightness temperature models for planets follow the Butler-JPL-Horizons 2012 framework, as explained in ALMA memo 594 \citep{butler12}. Specifically, the model for Uranus is based on the ESA4 model\footnote{https://www.cosmos.esa.int/web/herschel/calibrator-models}, and other SSOs are used with their flux scales referenced to Uranus within 5\% uncertainties \citep{kneissl23}. The secondary flux calibrators are automatically selected through online queries when the observations are carried out. As a result, the uncertainties originating from the absolute flux calibration are typically less than 5\% in Band~3 and 10\% in Bands~6 and 7 \citep[see][for more details]{remijan19}. 

For polarization analysis, we collected data reduced using the Analytic Matrix for ALMA POLArimetry (AMAPOLA\footnote{\url{https://www.alma.cl/~skameno/AMAPOLA/}}). This is a set of CASA-friendly python scripts to reduce the ALMA polarimetric data, as described in \citet{kameno23}. The dataset has been validated and used in previous studies of AGN jets such as M87 \citep{goddi21, goddi25} and OJ~287 \citep{jormanainen25}, as well as in studies of radio-quiet AGN \citep{shablovinskaia25}, supporting the reliability of the polarization measurements. 

\begin{table}[t]
    \centering
    \setlength{\tabcolsep}{6pt}
    \caption{Observation log for 3C~286.}
    \begin{tabular}{l l c c c}
    \toprule
     Instr. & Freq. & Period & $N_{\rm tot}$ & $N_{\rm pol}$ \\ 
     & $[$GHz$]$ & & & \\
     (1) & (2) & (3) & (4) & (5) \\  
    \midrule
    \multirow{4}{*}{ALMA} & 91.5 & 2018 May - 2024 Dec. & 207 & 330 \\
    & 103.5 & 2018 May - 2024 Dec. & 191 & 330 \\
    & 233.0 & 2018 May - 2024 Aug. & 32 & 73 \\
    & 343.4 & 2018 May - 2024 Dec. & 132 & 260 \\ 
    \midrule
    \multirow{2}{*}{IRAM} & 86 & 2006 Oct. - 2024 May & 106 & 106 \\ 
    & 229 & 2010 May - 2022 Jul. & 53 & 53 \\ 
    \midrule
    SMA & 225 & 2022 Jun. - 2025 Feb. & 56 & 56 \\ 
    \bottomrule
    \end{tabular}
    \raggedright 
    \tablecomments{Columns indicate: (1) instrument, (2) observing frequency, (3) monitoring period, (4) number of total flux measurements, and (5) number of polarization measurements. ‘IRAM’ refers to the IRAM 30-m telescope.}
   \label{tab:obs}
\end{table}\

\subsection{The IRAM 30-m Telescope} \label{subsec:Data_iram}
3C~286 has been monitored with the IRAM 30-m Telescope\footnote{In the following, the designation IRAM refers to its 30-m telescope.} at 86 and 229~GHz as part of the following three programs: the Polarimetric Monitoring of AGN at Millimeter Wavelengths (POLAMI) program \citep{agudo18_1, agudo18_3, thum18_2}, the Monitoring of AGN with Polarmietry at the IRAM 30-m telescope (MAPI) program \citep[e.g.,][]{agudo11_mapi}, and the polarimetric monitoring program of blazar OJ~287 \citep{agudo11_oj287}. It was typically observed with 15--20 other sources. The earlier data obtained from 2006 to 2014 suggested that both total flux and polarization of 3C~286 were stable at these frequencies during the monitoring period \citep{agudo18_1}. Here we present all the data obtained from October 2006 to May 2024 to extend the period of investigation. 

The monitoring of 3C~286 was performed with the XPOL polarimeter \citep{thum08} installed at the IRAM. The standard XPOL setup and calibration processes summarized in \citet{thum08} has been maintained throughout the monitoring \citep[e.g.,][]{agudo10, agudo12, agudo18_3, agudo18_1, thum18_2}. The data obtained before Spring 2009 were obtained using the orthogonal linearly polarized A100 and B100 heterodyne receivers only at 86~GHz. These receivers were then replaced with the new E090 and E230 pairs of orthogonal linearly polarized receivers, enabling us to observe at 86 and 229~GHz simultaneously after Spring 2009 \citep[see][for details]{thum08, agudo12}. The angular resolution of the telescope corresponds to 28" and 11" at 86 and 229~GHz, respectively. 

Each measurement includes a cross-scan pointing correction for each source, followed by amplitude, phase, and decorrelation loss calibration. After that, a set of wobbler switching on-off cycles with a total integration time of $\sim$8~minutes was conducted \citep{agudo10, agudo12}. The Kelvin-to-Jansky conversion factor, which converts antenna temperatures to flux densities, was derived based on the flux of Mars and Uranus measured from 2006 to 2014 \citep{agudo18_1}. The Mars model is based on the Lellouch \& Amri model\footnote{https://www.lesia.obspm.fr/perso/emmanuel-lellouch/mars/}, which provides the brightness temperature and flux of Mars for any frequency and date between 30-5000~GHz and year 1990-2036. For Uranus, the ESA4 model used for the Herschel Space Observatory mission \citep[e.g.,][]{bendo13, griffin13, orton14} was adopted. Consequently, the conversion factor was determined to be ($6.1\pm0.1$)~Jy K$^{-1}$ and ($8.6\pm0.4$)~Jy K$^{-1}$ for 86 and 229~GHz, respectively \citep{agudo18_1}. These factors were applied to the entire data obtained from 2006 to 2024\footnote{The dataset presented in our paper includes measurements previously published in \citet{agudo12} and \citet{agudo18_1}. Although different conversion factors were originally applied to those earlier data, we have reprocessed the entire dataset using the conversion factor from \citet{agudo18_1}, which is based on a larger number of observations, to ensure consistency.}. 

The instrumental polarization was measured during each session by observing unpolarized sources such as Mars or Uranus. Since the instrumental polarization and total power calibration parameters were sufficiently stable, they were interpolated for the few sessions where calibrators were not observed \citep[see][for details]{agudo18_1}. To cross-check the results of the polarization calibration, the Crab Nebula was observed \citep{agudo18_1}, as it is well-known for its stable and well-characterized polarization fraction and angle over a wide frequency range \citep[e.g.,][]{aumont10, weiland11, wiesemeyer11, ritacco18}. 

The final uncertainty for each total flux density measurement 
corresponds to a quadratic sum of the statistical uncertainties measured from the on-off cycles, the uncertainties in the planet models (conservatively considered as 5\%, which typically dominates the final uncertainty), and the uncertainties in the conversion factors \citep{agudo18_1}. This leads to average uncertainties in the total flux measurements of 3C~286 of 5.1\% at 86~GHz and 5.8\% at 229~GHz, respectively. 

\begin{figure*}[t!]
\centering 
\includegraphics[width=\textwidth]{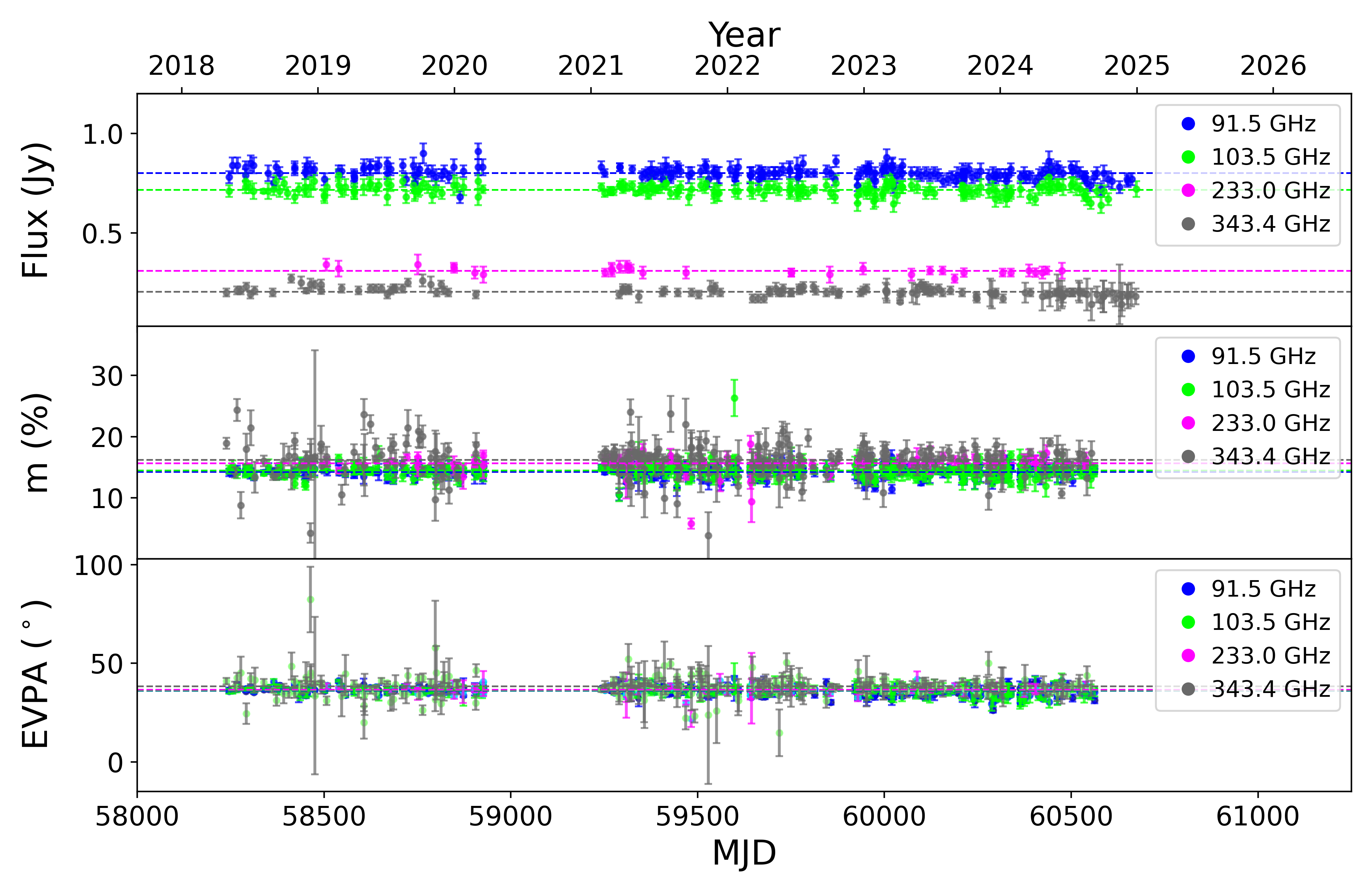}
\caption{Total flux (top), fractional polarization (middle), and EVPA (bottom) of 3C~286 obtained with ALMA at 91.5, 103.5, 233.0, and 343.4~GHz. The horizontal dotted lines represent the average values at each frequency. 
\label{fig:amapola_all}}
\end{figure*}

The uncertainties in the linear polarization fraction and EVPA of 3C~286 were derived from the statistical uncertainties in the on-off cycles and the dispersion of the Q and U Stokes parameters measured for Mars and Uranus, typically comparable to (or smaller than) 0.5\% for both Q and U Stokes parameters at both 86 and 229~GHz. This results in average uncertainties in the fractional polarization ($\langle \varepsilon_{m_\nu} \rangle$) of 0.9\% at 86~GHz and 2.8\% at 229~GHz, and average uncertainties in the EVPA ($\langle \varepsilon_{\chi_\nu} \rangle$) of 1.1$^\circ$ at 86~GHz and 5.1$^\circ$ at 229~GHz. We note that the uncertainty in the absolute EVPA measurement is less than 0.5$^\circ$ \citep{thum08, aumont10} and is therefore negligible compared to the uncertainties described above.

\subsection{SMA} \label{subsec:Data_sma}

The total flux of 3C~286 has been monitored with SMA over the past two decades\footnote{\url{http://sma1.sma.hawaii.edu/callist/callist.html}}. However, we do not include these measurements in our analysis due to inconsistencies in the flux calibration process and transitions from older solar system models to the CASA-standard models during the monitoring. These changes may introduce systematic offsets in the flux scale that are difficult to quantify reliably. Consequently, we only included recent data which were calibrated using a consistent process and model under the SMAPOL\footnote{\url{https://lweb.cfa.harvard.edu/sma/Newsletters/pdfFiles/SMA_NewsJan2024.pdf}} (SMA Monitoring of AGNs with POLarization) program.

SMAPOL is a program initiated in 2022 June to regularly monitor the polarization properties of 40 AGN jets that exhibit both prominent polarization and high-energy activities \citep{myserlis25}. 3C~286 has been monitored approximately every 2~weeks  to cross-check the polarization calibration. The observations were conducted in full polarization mode using the SMA polarization receiver system \citep{marrone06, marrone08} and SWARM correlator \citep{primiani16}. The separation between the center frequencies of the lower sideband (LSB) and upper sideband (USB) was set to about 20~GHz, centered at 225~GHz. The bandwidth of each sideband is 12~GHz, consisting of six chunks with a bandwidth of 2~GHz. 

During the observations, compact sources suitable for flux calibration were occasionally observed for 3--5 minutes. Their fluxes were calibrated mostly using SSOs such as Uranus, Neptune, Titan, Ganymede, and Callisto \citep{gurwell07}. As described in Section~\ref{subsec:Data_alma} and ~\ref{subsec:Data_iram}, the ESA4 model \citep{butler12} was also used to calibrate the flux of SSOs measured with the SMA. 

The initial data calibration such as flagging, bandpass, cross-receiver delays, complex gains, flux calibration, and RL-phase offset was performed using the MIR software package \footnote{\url{https://lweb.cfa.harvard.edu/~cqi/mircook.html}}, and the instrumental polarization was calibrated with the MIRIAD \citep{sault95}. The resulting fractional polarization and EVPA of 3C~286 have mean uncertainties of 2.8\% and 2.0$^\circ$, respectively.   

\begin{table*}[t]
    \centering
    \setlength{\tabcolsep}{5pt}
    \caption{Total flux measurements and best-fit parameters for 3C~286 using ALMA, IRAM, and SMA.}
    \begin{tabular}{c c c c c c c c c c}
    \toprule
     Instr. & $\nu$ & $S_\nu$ & $\sigma_{S_\nu}$ & $\langle \varepsilon_{S_\nu} \rangle$ & c$_{\rm 0}$ & c$_{\rm 1}$ & $\Delta S_{\nu}$ & $R$ \\  
    & $[$GHz$]$ & [Jy] & [Jy] & [Jy] & [$\times 10^{-5}$~Jy/day] & [Jy] & [Jy] & [\%] & \\ 
    & & (1) & (2) & (3) & (4) & (5) & (6) & (7) \\ 
    \midrule
    \multirow{4}{*}{ALMA} & 91.5 & 0.802 & 0.029 & 0.033 & $-1.16\pm0.31$ & $0.820\pm0.006$ & $-0.028$ & $-4$ \\
    & 103.5 & 0.717 & 0.025 & 0.032 & $-0.66\pm0.30$ & $0.726\pm0.005$ & $-0.016$ & $-2$ \\
    & 233.0 & 0.309 & 0.016 & 0.027 & $-1.46\pm0.73$ & $0.333\pm0.013$ & $-0.037$ & $-11$ \\
    & 343.4 & 0.204 & 0.021 & 0.031 & $-1.33\pm0.27$ & $0.224\pm0.005$ & $-0.033$ & $-15$ \\ 
    \midrule 
    \multirow{2}{*}{IRAM} & 86 & 0.900 & 0.047 & 0.046 & $-0.33\pm0.26$ & $0.890\pm0.004$ & $-0.021$ & $-2$ \\ 
     & 229 & 0.328 & 0.047 & 0.019 & $0.58\pm0.18$ & $0.325\pm0.003$ & $+0.027$ & $+9$ \\ 
     \midrule 
     SMA & 225 & 0.322 & 0.052 & 0.039 & - & - & - & - \\ 
    \bottomrule
    \end{tabular}
    \raggedright 
    \tablecomments{Columns indicate: (1) unweighted mean total flux, (2) the standard deviation of the total flux, (3) the average measurement uncertainty, (4)--(5)  the best-fit parameters for Equation~\ref{eq:fit}, (6) the amplitude of the flux decrease during the monitoring period, and (7) the percentage flux change observed during the monitoring period, as determined from the best-fit parameters. Across all frequencies, the median total flux values agree with the mean values within 1--2\%. 
} 
    \label{tab:flux_summary}
\end{table*}

\subsection{VLA}

The stability of 3C~286 at 1--50~GHz allows us to investigate the spectral flux density across centimeter to millimeter/submillimeter wavelengths by utilizing historical measurements at centimeter wavelengths. For this comparison, we adopted the total flux measurements using VLA at 1--50~GHz \citep{perley13_flux}, along with polarization measurements from the same instrument. Specifically, we took the EVPA measurements at 5, 15, 25, 33, and 45 GHz from Table~2 in \citet{perley13_pol}, representing the weighted mean of measurements obtained at each frequency from 1995 to 2012. The fractional polarization values were taken from Table~5 of \citet{perley13_pol}, corresponding to measurements at 1.5, 4.9, 8.4, 15.0, 22.5, and 43.3~GHz for epoch 2000.0 in the same reference. 

\begin{figure}[t!]
\centering 
\includegraphics[width=\columnwidth]{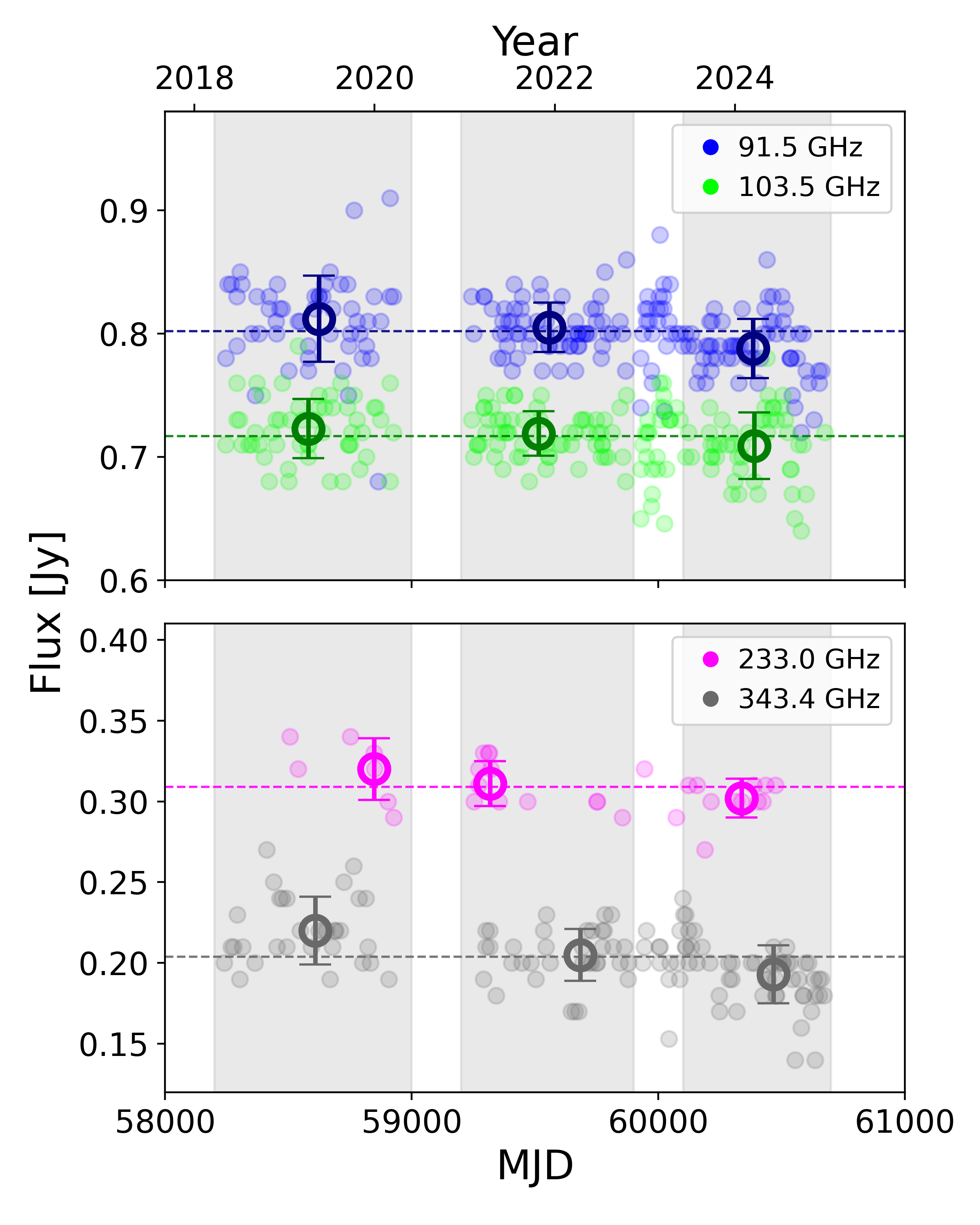}
\caption{The total flux of 3C~286 measured with ALMA in Band~3 (top) and Bands 6 and 7 (bottom). Filled circles without error bars represent the full dataset. Larger highlighted circles with error bars indicate the average flux and the standard deviation at each frequency (Table~\ref{tab:app_flux}), displayed at the median date of the corresponding period: Period I (MJD 58200--59000), Period II (MJD 59200--59900), and Period III (MJD $>$ 60100). Shaded regions correspond to these three periods. Dashed horizontal lines indicate the mean flux values over the entire monitoring period. 
\label{fig:alma_flux}}
\end{figure}

\begin{figure*}[t!]
\centering 
\includegraphics[width=\textwidth]{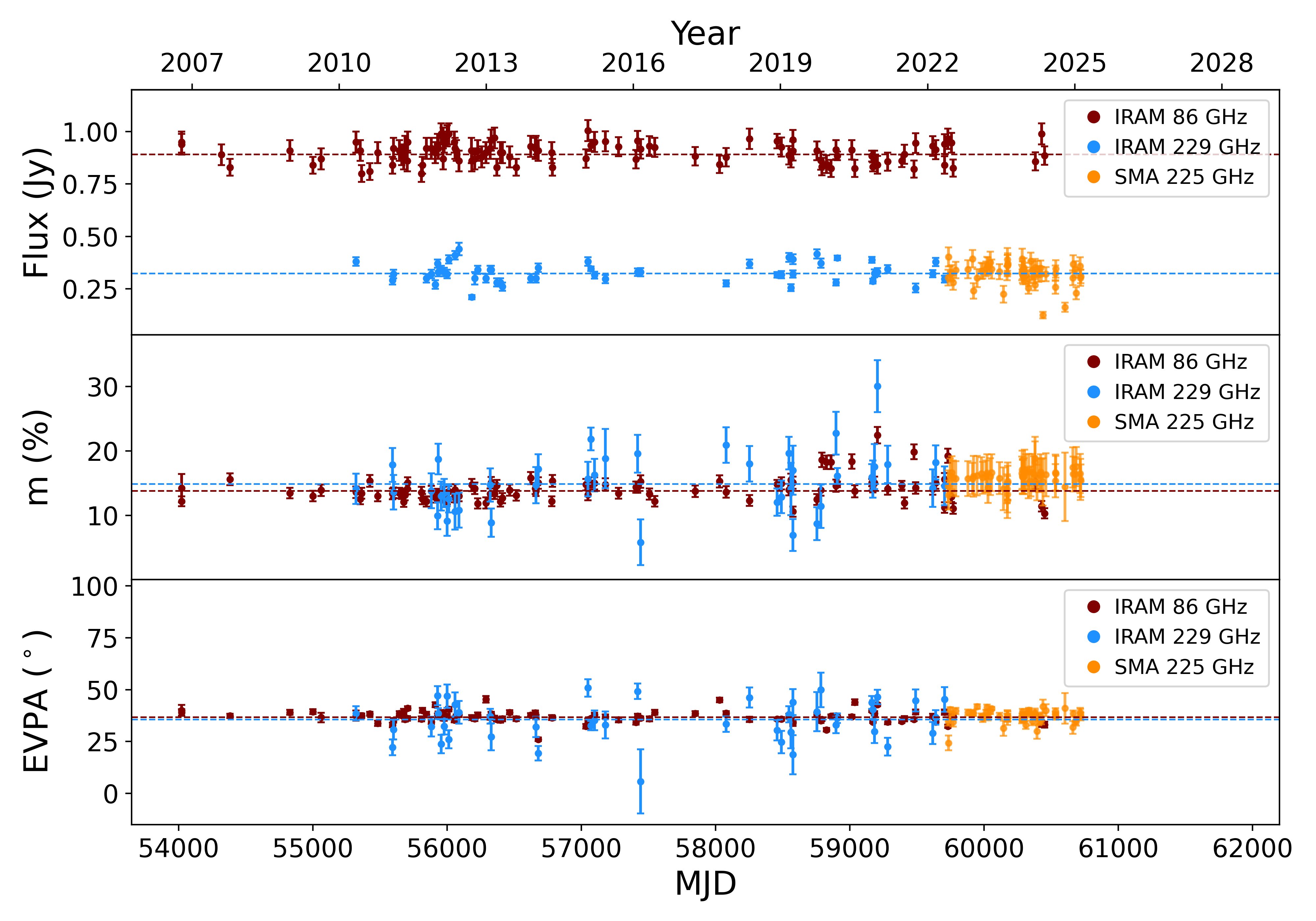}
\caption{Total flux (top), fractional polarization (middle), and EVPA (bottom) of 3C~286 obtained with IRAM at 86 and 229~GHz and SMA at 225~GHz. The horizontal dotted lines represent the average values at each frequency. Note that the y-axis scales are identical to those in Figure~\ref{fig:amapola_all}.
\label{fig:iram_all}}
\end{figure*}

\section{Results} \label{sec:analysis}

\subsection{Total flux}\label{sec:result_total}
In Figure~\ref{fig:amapola_all}, we present the total flux, fractional polarization, and EVPA of 3C~286 measured with ALMA at four frequencies: 91.5, 103.5, 233.0, and 343.4~GHz. Notably, the total flux appears to gradually decrease over time, exhibiting fluctuations on short timescales. For example, the total flux at 91.5~GHz obtained before MJD~59000 lies mostly above the total flux averaged over the entire monitoring period. In contrast, the total flux measured after MJD~60000 falls generally below this average, suggesting a decreasing trend.  

To investigate this in further detail, we divided the ALMA data into three groups: the first, obtained between MJD~58230 and 59000, referred to as Period I, the second, obtained between MJD~59200 and 59900, referred to as Period II, and the third, after MJD~60100, referred to as Period III. Indeed, the average total flux values decrease at all frequencies from Period~I to Period~III (Figure~\ref{fig:alma_flux}), albeit the magnitude of the decline is comparable to the standard deviations of each period (see Table~\ref{tab:app_flux} in Appendix~\ref{sec:appendix_flux}).

\begin{figure}[t!]
\centering 
\includegraphics[width=\columnwidth]{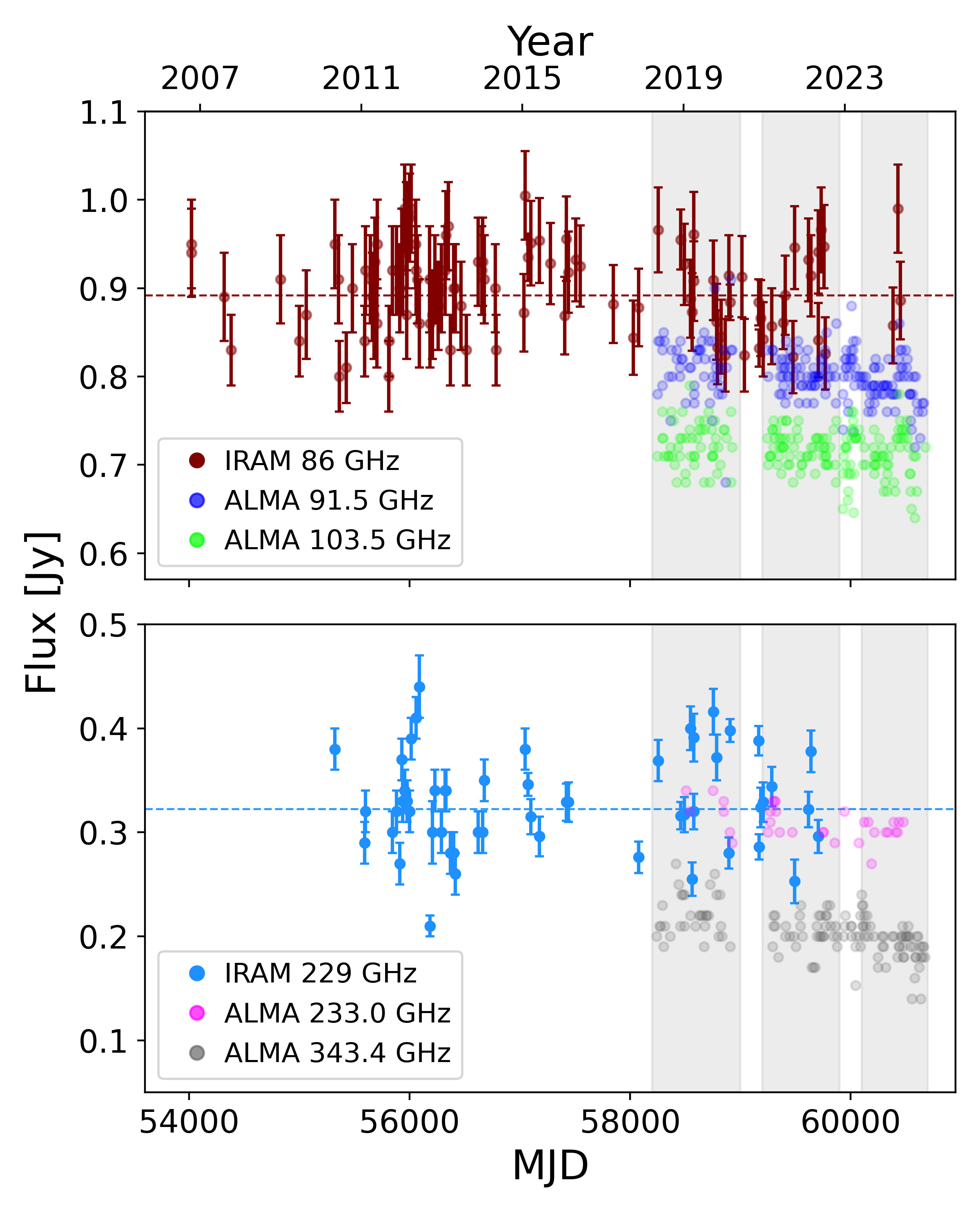}
\caption{Total flux of 3C~286 measured with IRAM at 86~GHz and ALMA in Band 3 (top), and with IRAM at 229~GHz and ALMA in Bands 6 and 7 (bottom). Dashed horizontal lines indicate the mean flux values measured with IRAM over its monitoring period for each frequency. Shaded regions correspond to these three periods shown in Figure~\ref{fig:alma_flux}. 
\label{fig:alma_iram_flux}}
\end{figure}

To further investigate the long-term trend in the ALMA data, we fitted all the total flux measurements at each frequency with the following function: 
\begin{equation}\label{eq:fit}
    S_{\nu} = c_0 \times (t-58000) + c_1
\end{equation}
\noindent where $S_{\nu}$ is the total flux in Jy, and $t$ is the observation date in MJD. The best-fit parameters are summarized in Table~\ref{tab:flux_summary}, indicating that the flux has decreased by 2\%--15\% over the ALMA monitoring period. The decrease of 4\% at 91.5~GHz and 2\% at 103.5~GHz might not be significant given the typical measurement uncertainty of 5\% in Band~3 \citep{remijan19}. However, the declines of 11\% at 233.0~GHz and 15\% at 343.4~GHz exceed the typical uncertainties of 10\% in Bands~6~and~7. 

In contrast, the total flux measurements obtained with IRAM do not exhibit any clear decreasing trend at both 86 and 229 GHz (Figure~\ref{fig:iram_all}). 
The best-fit parameters derived from fitting the IRAM data at 86~GHz to Equation~\ref{eq:fit} suggest that the secular flux change from 2006 to 2024 ($\Delta S_{\nu}$) is smaller than the change observed in the ALMA data at 91.5~GHz from 2018 to 2024 (Table~\ref{tab:flux_summary}). Furthermore, the flux at 91.5~GHz expected from the spectral index derived using IRAM measurements at 86 and 229~GHz (see Section~\ref{sec:cooling}) is 0.84~Jy. This is 5\% larger than the 0.80~Jy measured with ALMA at the same frequency (see also Figure~\ref{fig:alma_iram_flux}).   

At 233\footnote{We do not use the SMA data to test the flux stability of 3C~286, as its monitoring period from June 2022 to February 2025 is too short for a reliable investigation.} and 343.4~GHz, the decrease in the total flux ($\Delta S_{\rm \nu}$) measured with ALMA from 2018 to 2024 exceed their standard deviations and measurement uncertainties (Table~\ref{tab:flux_summary}). In contrast, no such decreasing trend is identified in the IRAM data at 229~GHz over the longer period from 2010 to 2022 \citep[see also][]{agudo12, agudo18_1}. One explanation to consider for the discrepancy between the ALMA and IRAM data would be the increasing brightness of Uranus \citep[e.g.,][]{kneissl23}, which has been used as the primary flux calibrator for ALMA.

\begin{figure*}[t!]
\centering 
\includegraphics[width=0.85\textwidth]{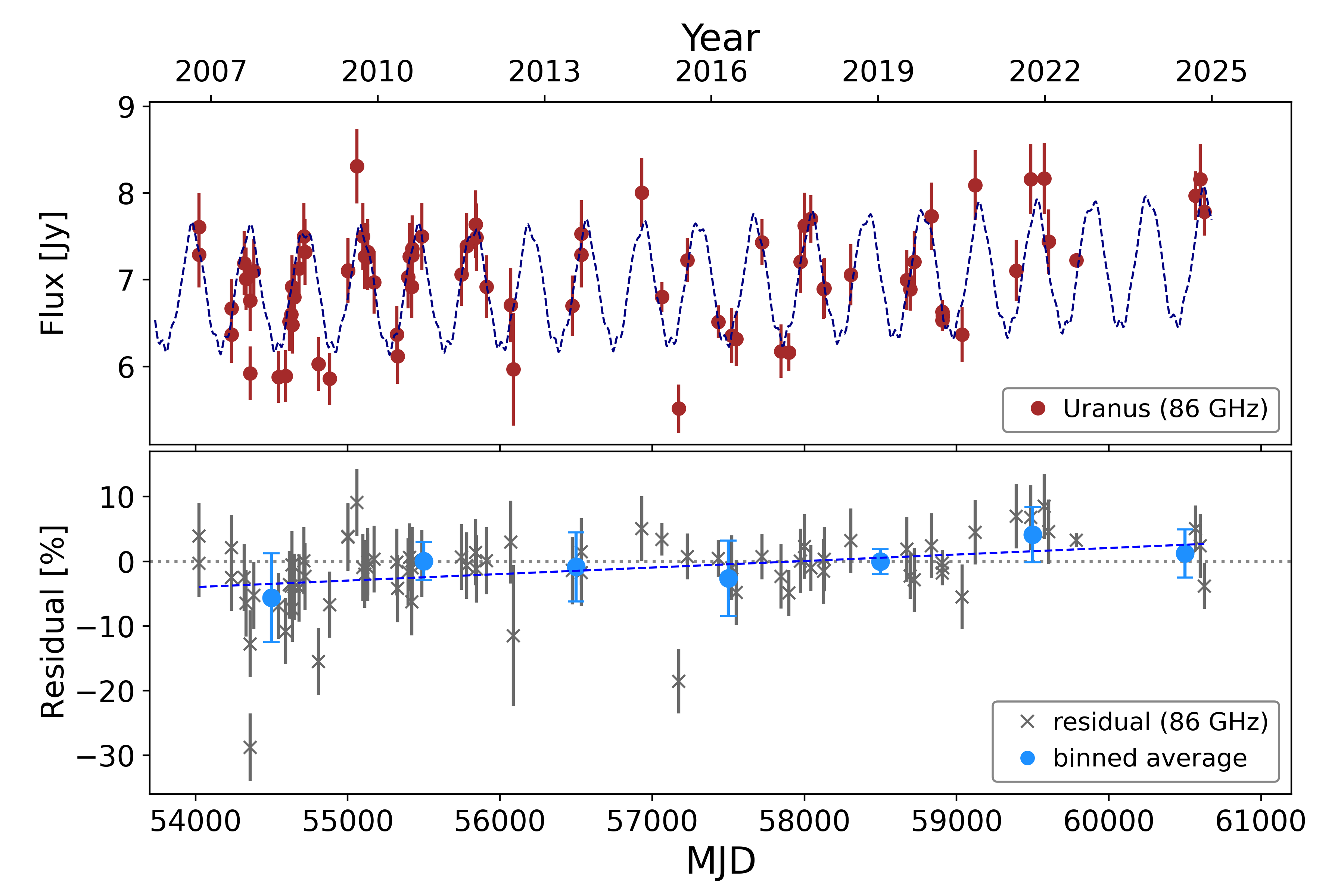}
\caption{Top: The total flux of Uranus measured with IRAM at 86~GHz. The blue dashed curve represents the ESA4 model. Bottom: The residual flux, i.e., the excess flux relative to the model normalized by the observed flux, is shown with cross marks. The light blue points represent the average residual flux in 1000-day bins. The blue dashed line represents the best-fit lines to the entire residual flux. The ESA4 model appears to agree best with the data around 2017. 
\label{fig:uranus}}
\end{figure*}

Thanks to its brightness and relatively small angular size of $\sim$3~arcseconds, Uranus has been used as a primary flux calibrator at millimeter/submillimeter wavelengths based on the ESA4 model, which assumes its absolute brightness is constant over time. However, its large obliquity of 82$^\circ$ causes the sub-Earth latitude to vary over time, leading us to see Uranus' polar and equatorial regions alternately. Notably, the brightness temperatures of Uranus' north and south poles are higher than that of its equatorial region \citep[e.g.,][]{briggs80, jaffe84, depater88, berge88, molter21, akins23}. This implies that the brightness temperature of Uranus would alternate between its maximum and minimum every 21~years, given its 84-year orbital period around the Sun (e.g., see Figure~9 in \citet{fletcher21}). This variation is expected to be further amplified by the ring around Uranus, which becomes brighter when the polar regions are oriented toward Earth \citep{butler12}. 

This has been confirmed through long-term monitoring observations. The last solstice during which the south pole of Uranus was toward Earth was in 1986, while the last equinox was in 2007. Accordingly, the brightness of Uranus increased from the mid-1960s to mid-1980s, then decreased until the mid-2000s, showing fluctuations on short timescales \citep[e.g.,][]{klein78, klein06, perley13_flux, kramer08}. 
After the last equinox in 2007, the north pole of Uranus has been rotating into our view and will be fully oriented toward Earth by the next solstice in 2028. Accordingly, we expect that the brightness of Uranus has been increasing since 2007. 

Figure~\ref{fig:uranus} shows the flux of Uranus monitored with IRAM at 86~GHz since 2006. It exhibits periodic increase and decrease as expected from the seasonal variation in the distance between Earth and Uranus. If the intrinsic brightness of Uranus has indeed increased, the residual flux, defined as the excess flux relative to the model, should have been rising during the monitoring. Notably, the residual flux binned over 1000-day intervals shows an increasing trend\footnote{This trend is the inverse of that observed from 1985 to 2005, when the brightness of Uranus was decreasing from solstice to equinox \citep[e.g., Figure~2 in][]{kramer08}.}. A fit using Equation~\ref{eq:fit} to the entire residual flux indicates a systematic increase of 6.6\%$\pm$1.5\% from 2006 to 2024, albeit the magnitude of the increase is comparable to the average uncertainty in the residual flux of 4.8\%. 
Furthermore, the ratio of the brightness of Uranus to Neptune, monitored with ALMA from 2018 to 2024, also suggests that the brightness of Uranus has increased by 4\% during the monitoring period\footnote{However, we also note that the flux scaling factors derived from Uranus and Neptune as seen by ALMA is in agreement within 2\% since 2022 \citep{kneissl23}.} \citep{kneissl23}. These results suggest that the ESA4 model, which assumes constant brightness for Uranus, needs to be modified. This, in turn, suggests that brightness variations in Uranus could partially account for the apparent decrease in the flux of 3C~286 measured with ALMA.

In fact, Uranus has not been observed together with 3C~286, as their Right Ascensions are nearly opposite. This indicates that the brightness of Uranus may not have been directly used to scale the flux of 3C~286. However, even when no Uranus data were used to directly calibrate the flux of 3C~286, any flux correction factors derived from other SSOs that deviate by more than 5\% from those derived from Uranus were excluded from the flux calibration process \citep{kneissl23}. Consequently, only SSOs flux measurements that provided correction factors consistent with those from Uranus within 5\% were used. 

In contrast, IRAM did not use the recent flux of Uranus for flux calibration. As explained in Section~\ref{subsec:Data_iram}, the Kelvin-to-Jansky conversion factor for IRAM was estimated using observations of Mars and Uranus between 2006 and 2014. The observed fluxes of these planets were scaled to their models: the Lellouch \& Amri model for Mars and the ESA4 model for Uranus \citep{butler12}. Although the IRAM data also adopted the ESA4 model, it was used for the data obtained from 2006 to 2014 when Uranus was relatively close to equinox in 2007, and were applied to the entire data. As a result, the IRAM data depend only weakly on the fluxes of Uranus\footnote{Given that the ESA4 model appears to agree best with the data around 2017 (Figure~\ref{fig:uranus}), the IRAM flux measurements could have been underestimated by $\sim2\%$ at 86~GHz, although their stability remains intact.}. 

Nevertheless, the change in the brightness of Uranus alone cannot fully account for the decreasing trend observed in the ALMA measurements, as the magnitudes of the decrease in Band~6 and 7 exceed 10\%. 
This implies that additional factors may contribute to the discrepancy, such as non-linear atmospheric variations of Uranus \citep[e.g.,][]{orton86, depater89, griffin93} or instrumental effects. These issues will be investigated further in future studies.

\begin{table*}[t!]
    \centering
    \setlength{\tabcolsep}{3.3pt}
    \caption{Polarization measurements and best-fit parameters for 3C~286.}
    \begin{tabular}{c c c c c c c c c c c c c c c}
    \toprule
    & & \multicolumn{6}{c}{EVPA} & & \multicolumn{6}{c}{Fractional polarization} \\ 
    \cmidrule{3-8}
    \cmidrule{10-15}
     Instr. & $\nu$ & $\chi_\nu$ & $\sigma_{\chi_\nu}$ & $\langle \varepsilon_{\chi_\nu} \rangle$ & c$_{\rm 0}$ & c$_{\rm 1}$ & $|\Delta\chi|$ & & $m_\nu$ & $\sigma_{m_\nu}$ & $\langle \varepsilon_{m_\nu} \rangle$ & c$_{\rm 0}$ & c$_{\rm 1}$ & $|\Delta m|$ \\ 
    & $[$GHz$]$ & [$^\circ$] & [$^\circ$] & [$^\circ$] & [$\times 10^{-4}\ ^\circ$/day] & [$^\circ$]  & [$^\circ$] 
    & & [\%] & [\%] & [\%] & [$\times 10^{-5}\ \%$/day] & [\%]  & [\%] \\ 
    & & (1) & (2) & (3) & (4) & (5) & (6) & & (1) & (2) & (3) & (4) & (5) & (6) \\  
    \midrule
    \multirow{5}{*}{ALMA} & 91.5 & 36.2 & 2.9 & 2.4 & $-7.20\pm0.48$ & $38.23\pm0.10$ & $2.5$ 
    & & 14.3 & 0.9 & 0.8 & $7.85\pm4.56$ & $14.14\pm0.11$ & $0.3$ \\
    & 103.5 & 36.3 & 1.5 & 1.2 & $-4.37\pm0.57$ & $37.48\pm0.12$ & $1.5$ 
    & & 14.5 & 1.1 & 0.9 & $-3.75\pm5.04$ & $14.58\pm0.12$ & $0.1$ \\
    & 233.0 & 36.7 & 2.7 & 2.0 & $-4.89\pm1.24$ & $38.24\pm0.28$ & $1.7$ 
    & & 15.7 & 1.8 & 1.0 & $-10.02\pm10.24$ & $16.30\pm0.25$ & $0.3$ \\
    & 343.4 & 38.3 & 5.5 & 3.8 & $-1.58\pm0.80$ & $39.25\pm0.19$ & $0.5$ 
    & & 16.2 & 2.8 & 1.8 & $-24.15\pm9.20$ & $16.94\pm0.23$ & $0.8$ \\ 
    \midrule 
    \multirow{2}{*}{IRAM} & 86 & 37.0 & 2.7 & 1.1 & $-1.55\pm0.66$ & $36.93\pm0.10$ & $1.0$ & & 14.0 & 1.9 & 0.9 & $22.36\pm4.99$ & $13.77\pm0.08$ & $1.4$  \\ 
     & 229 & 35.0 & 9.6 & 5.1 & $10.65\pm4.45$ & $35.20\pm0.10$ & $4.9$ & & 14.5 & 5.4 & 2.8 & $74.26\pm27.09$ & $14.55\pm0.40$ & $3.4$ \\  
    \midrule 
    SMA & 225 & 37.2 & 3.0 & 2.0 & - & - & - & & 15.9 & 1.2 & 2.8 & - & - & - \\  
    \bottomrule
    \end{tabular}
    \raggedright 
    \tablecomments{The columns indicate (1) unweighted mean, (2) standard deviation, (3) average of the measurement uncertainties, (4)--(5) the best-fit parameters derived using Equation~\ref{eq:fit}, and (6) the changes observed during the monitoring period for each instrument at each frequency, as determined from the best-fit parameters. Across all frequencies, the median EVPA and fractional polarization agree with the mean values within 0.4$^\circ$ and 0.3\%, respectively. 
} 
    \label{tab:pol_summary}
\end{table*}

In summary, the brightness of Uranus has likely been increasing since 2007 possibly due to the gradual rotation of its north pole into our view. 
This could partially contribute to the apparent decrease in the flux of 3C~286 measured with ALMA, for which Uranus was used as the primary flux calibrator. Taking these factors into account, we conclude that the total flux of 3C~286 is nearly constant at frequencies up to 229~GHz throughout the 20-year monitoring period (Table~\ref{tab:flux_summary}). In addition, these results suggest that the ESA4 model may need to be revised to account for the brightness variation in Uranus caused by its pole rotation.

Regarding the ALMA flux calibration, a flux decrease of 3C~286 of up to 4\% could be attributed to this brightening of Uranus, as already noted by \citet{kneissl23}. However, the additional decrease observed in Band~6 and 7 may arise from instrumental effects, short-term variations intrinsic to Uranus and 3C~286, or a combination of both. Continued monitoring of both 3C~286 and Uranus along with other SSOs will help clarify the origin of the flux decrease.

\subsection{Polarization}\label{sec:result_pol}

In contrast to the total flux, no secular changes are observed in the fractional polarization and EVPA measured with ALMA (Figure~\ref{fig:amapola_all}). We compared the average fractional polarization and EVPA values across Period I--III, as we did for the total flux to investigate their variability. The results show that the fractional polarization and EVPA values remain constant across Period I---III, within their standard deviations and average measurement uncertainties (see Table~\ref{tab:app_alma_evpa} and \ref{tab:app_alma_fracpol} in Appendix~\ref{sec:appendix_flux}). Furthermore, the best-fit parameters derived using Equation~\ref{eq:fit} indicate that any potential secular changes in both EVPA and fractional polarization during the monitoring period are smaller than or comparable to their standard deviations and/or uncertainties at all frequencies (Table~\ref{tab:pol_summary}). This suggests that there are no significant trends exceeding the measurement uncertainties. 

This finding is consistent with the IRAM monitoring over a longer period. The potential secular variations in the IRAM data, also derived using the best-fit parameters to Equation~\ref{eq:fit}, are likewise smaller than or comparable to their standard deviations and/or uncertainties at both 86 and 229~GHz (Table~\ref{tab:pol_summary}). 

Likewise, long-term monitoring observations with ALMA and IRAM suggest that both EVPA and fractional polarization of 3C~286 have remained nearly constant over the last $\sim$20~years. This stability allows us to obtain the fractional polarization and EVPA values with unprecedented accuracy at millimeter/submillimeter wavelengths by stacking all the data. Therefore, we provide the weighted mean of the EVPA and fractional polarization, as well as the total flux, as the reference values for 3C~286 (Table~\ref{tab:pol_reference}).

\begin{figure*}[t!]
\centering 
\includegraphics[scale=0.6]{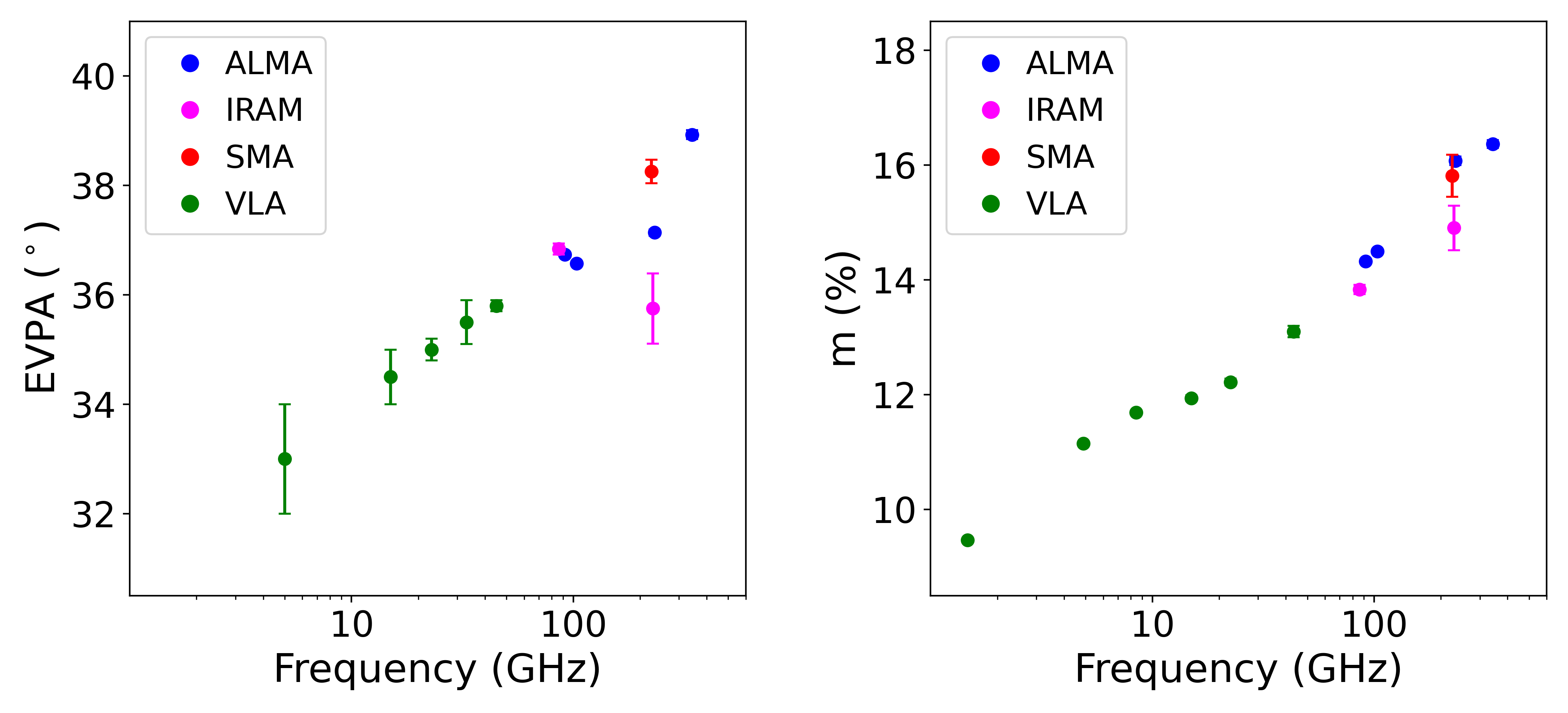}
\caption{The reference values for the EVPA and fractional polarization of 3C~286 obtained with ALMA, IRAM, SMA (this work), and VLA \citep{perley13_pol} in the observing frame. Blue, magenta, red, and green colors represent the ALMA, IRAM, SMA, and VLA data, respectively. 
\label{fig:bothpol}}
\end{figure*}

In Figure~\ref{fig:bothpol}, we present the reference values for the EVPA and fractional polarization of 3C~286, along with the values measured with VLA at 5--45~GHz \citep{perley13_pol}. Notably, both quantities commonly exhibit a gradual change as a function of frequency: the EVPA gradually rotates, while the fractional polarization increases across the range from 5 to 343.4~GHz. 
These trends, especially at millimeter/submillimeter, were unclear in the past due to insufficient monitoring periods and limited number of measurements, which hindered testing stability and prevented achieving the signal-to-noise ratio required for robust analysis \citep[see Figure~3 in][]{hull16}. Our analysis with the ALMA, IRAM, and SMA data have revealed that the trends in both fractional polarization and EVPA indeed extends up to 343.4~GHz. 

\begin{table}[t!]
    \centering
    \setlength{\tabcolsep}{4pt}
    \caption{Reference values for 3C~286}
    \begin{tabular}{c c c c c}
    \toprule
    Instr. & $\nu$ [GHz] & S$_{\nu}$ [Jy] & $\chi$ [$^\circ$] & $m$ [\%] \\ 
    \midrule
    \multirow{4}{*}{ALMA} & 91.5 & $-$ & $36.74\pm0.03$ & $14.32\pm0.03$ \\
    & 103.5 & $-$& $36.57\pm0.04$ & $14.50\pm0.04$ \\
    & 233.0 & $-$& $37.14\pm0.06$ & $16.07\pm0.07$ \\
    & 343.4 & $-$& $38.93\pm0.08$ & $16.37\pm0.07$ \\ 
    \midrule 
    \multirow{2}{*}{IRAM} & 86 & $0.892\pm0.004$ & $36.84\pm0.10$ & $13.83\pm0.08$ \\ 
    & 229 & $0.323\pm0.003$ & $35.75\pm0.64$ & $14.48\pm0.38$ \\ 
    \midrule 
    SMA & 225 & - & $38.26\pm0.22$ & $15.81\pm0.37$ \\ 
    \bottomrule
    \end{tabular}
    \raggedright
    \tablecomments{The weighted mean values and their standard uncertainties for the total flux, EVPA, and fractional polarization measurements. We provide reference values for the total flux based solely on the IRAM data, where flux stability is confirmed.} 
    \label{tab:pol_reference}
\end{table}

We note that the reference EVPA from IRAM at 229~GHz appears to deviate from the trend observed in the reference EVPA values obtained with other instruments. This is probably because the EVPA uncertainty for IRAM is larger than that for ALMA and SMA. As shown in Table~\ref{tab:flux_summary} and Figure~\ref{fig:iram_all}, the EVPA measurements with IRAM at 229~GHz shows much larger scatter compared to the other instruments. Moreover, the average of the EVPA uncertainties for IRAM at 229~GHz is 5.1$^\circ$, while the standard deviation of those measurements is 9.6$^\circ$. This discrepancy suggests that the EVPA uncertainties for IRAM may have been underestimated. Since the uncertainties in Figure~\ref{fig:bothpol} are based on those uncertainties, the actual EVPA uncertainty for IRAM would be larger than what is shown.

\section{Discussion} \label{sec:Discussions}

\subsection{Synchrotron cooling} \label{sec:cooling}
\begin{figure}[t!]
\centering 
\includegraphics[width=\columnwidth]{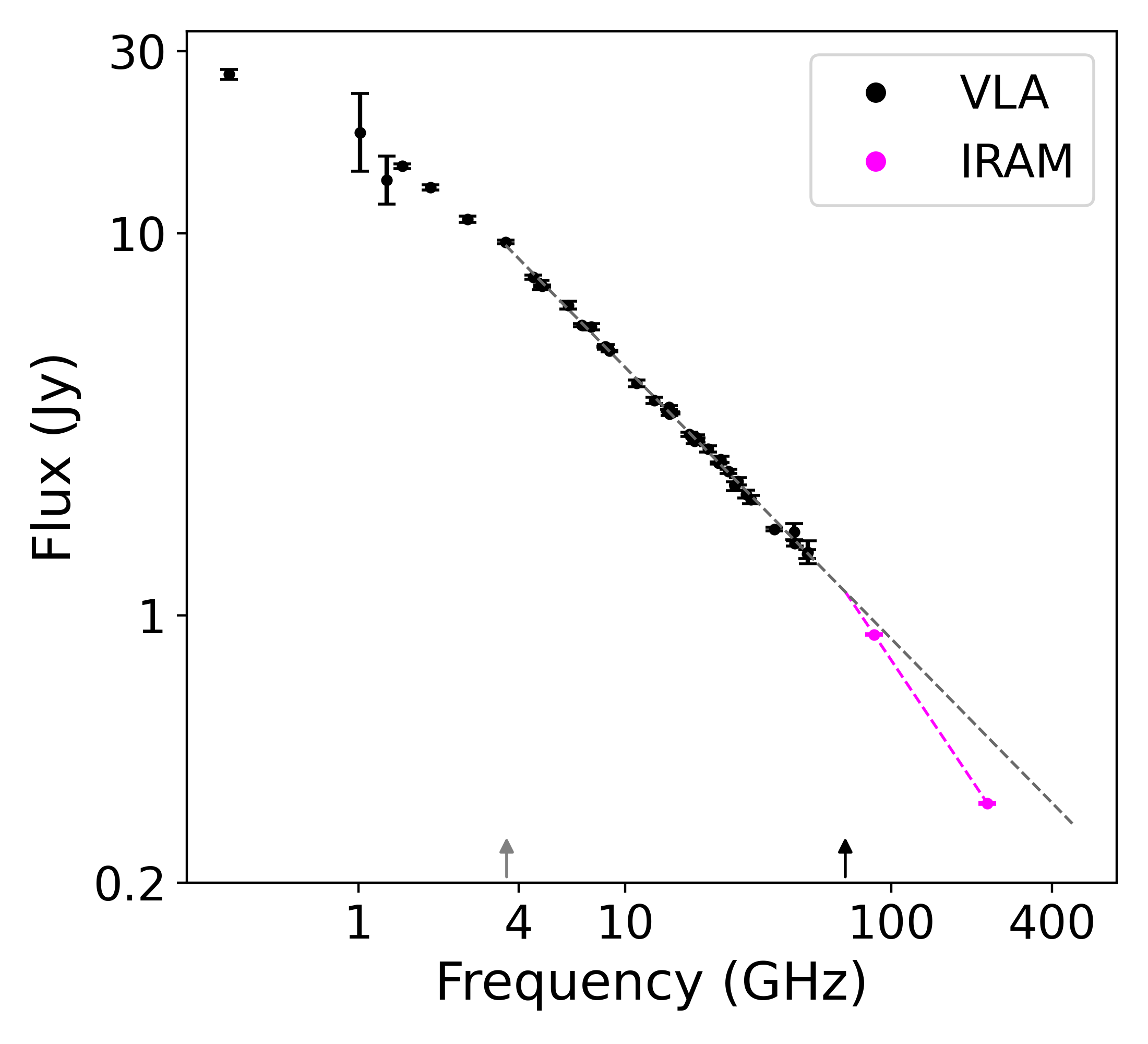}
\caption{The spectral flux density of 3C~286 between 1 and 229~GHz. The black and magenta data points show the total flux measured with VLA \citep{perley13_flux} and IRAM, respectively. The gray and magenta dashed lines show the best-fit to the total flux measured with VLA and IRAM, respectively. The black and gray arrows on the x-axis indicate the break frequency at 67.2 GHz and a potential break frequency at 3.6~GHz (see Section~\ref{sec:b_strength}), respectively.
\label{fig:spectral}}
\end{figure}

The stability of 3C~286 across centimeter to millimeter/submillimeter wavelengths enables a detailed investigation of its spectrum by combining data from centimeter to millimeter wavelengths. For the frequencies higher than those covered by VLA, we use only the IRAM measurements to determine the spectral index, as the ALMA flux appears to vary with different rates in different frequency bands. The spectrum steepens at higher frequencies, resulting in a spectral break where the spectral index~$\alpha$ (defined by $S_{\nu}\propto \nu^{-\alpha}$) changes between lower and higher frequency ranges (Figure~\ref{fig:spectral}). Specifically, the spectral index between 3 and 50~GHz, where the flux nearly follows a simple power-law, yields a spectral index of $\alpha=0.712\pm0.003$. In contrast, the spectral index between 86 and 229~GHz is $\alpha=1.039\pm0.008$  (Table~\ref{tab:comp_spectral}). This difference in spectral index between the two frequency ranges leads to a spectral break at $67.2\pm1.2$~GHz\footnote{As described in Section~\ref{sec:result_total}, if the IRAM flux measurements were underestimated by $\sim2\%$, the break frequency is shifted to 71.4~GHz.}. 

Such a spectral break can be attributed to the process of synchrotron cooling (or aging), which refers to the process by which the spectrum of synchrotron radiation changes over time as the electrons lose energy via radiation \citep[e.g.,][]{kardashev62, pacholczyk70, jaffe73}. This process results in a steepening of the synchrotron spectrum at high frequencies, as higher-energy electrons lose energy more rapidly than lower-energy electrons. For example, a relativistic electron with total energy $E=\gamma m_e c^2$ moving in a magnetic field radiates energy at a rate proportional to the square of its energy \citep{rybicki79}

\begin{equation}
    P=\frac{dE}{dt}\propto E^2 B^2
\end{equation}

\noindent where $E$ is the total energy of the electron and $B$ is the magnetic field strength. This suggests that higher-energy electrons lose their energy more rapidly when the magnetic field strength is constant over time. In addition, an electron emits most of its energy near the critical frequency $\nu_{\rm c}$, above which the synchrotron spectrum fades away, and which is proportional to the square of its energy \citep{rybicki79}

\begin{equation}
    \nu_c \propto E^2 B.
\end{equation}

\noindent This indicates that higher-energy electrons lose most of their energy at higher frequencies \citep{condon16}. Likewise, the fact that both the energy loss rate and critical frequency are proportional to $E^2$ suggests that the depletion of high-energy electrons steepens the spectrum at these higher frequencies more rapidly. This scenario aligns well with the spectral steepening at millimeter/submillimeter wavelengths observed in 3C~286.

\begin{table}[t]
    \centering
    \setlength{\tabcolsep}{4pt}
    \caption{Spectral index and break frequency.}
    \begin{tabular}{c c c c}
    \toprule
    $\nu_{1}$ -- $\nu_{2}$ $[$GHz$]$ & $\alpha$ & $\nu_{\rm br}$ [GHz] & S$_{\rm br}$ [Jy]\\ 
    (1) & (2) & (3) & (4) \\ 
    \midrule
    3--50 & $0.712\pm0.003$ & \multirow{2}{*}{$67.2\pm1.2$} & \multirow{2}{*}{$1.2\pm0.1$} \\ 
    86--229 & $1.039\pm0.008$ & & \\ 
    \bottomrule
    \end{tabular}
    \raggedright 
    \tablecomments{The columns indicate (1) frequency range, (2) the spectral index, (3) the break frequency, and (4) the total flux at the break frequency in each period. } 
    \label{tab:comp_spectral}
\end{table}

Such spectral steepening is widely observed in various types of synchrotron emitting sources, including the Active Galactic Nuclei (AGN) jets classified as Fanaroff \& Riley class II \citep[e.g.,][]{fr74, myers85, alexander87, machalski09, machalski10, harwood13}, the Compact Steep Spectrum (CSS) sources \citep[e.g.,][]{murgia99, murgia03, orienti04}, pulsar wind nebula, and supernova remnants \citep[e.g.,][]{reynolds84, bock05, reynolds09, bucciantini11, lyutikov19}. The spectral steepening in 3C~286 also aligns well with the characteristics of CSS sources, the class to which 3C~286 belongs.

\subsection{Synchrotron age and magnetic field strength}\label{sec:b_strength}

The presence in the data of the break frequency $\nu_{\rm br}$ allows us to estimate the age of the source. Assuming (i) a \textit{continuous injection model}, where new relativistic electrons with a power-law energy distribution are continuously injected to the synchrotron-emitting source\footnote{The standard shape of the synchrotron spectrum below the critical frequency is well explained by the continuous injection of relativistic electrons \citep{kardashev62, murgia99}.}, (ii) no expansion of the emitting source, and (iii) constant magnetic field strength, the time elapsed after the formation of the emitting source, $\tau_{\rm syn}$ (in Myrs), can be expressed as \citep{murgia99, murgia03}:

\begin{equation}\label{eq:tau_syn}
    \tau_{\rm syn} = 1610 \frac{B_{\rm syn}^{0.5}}{B_{\rm syn}^2+B_{\rm CMB}^2}\frac{1}{[\nu_{\rm br}(1+z)]^{0.5}}
\end{equation}

\noindent where $B_{\rm syn}$ is the average magnetic field strength in the emitting region (in $\mu$G), $B_{\rm CMB}=3.25(1+z)^2$ is the magnetic field strength equivalent to the cosmic microwave background (in $\mu$G), and $\nu_{\rm br}$ is the break frequency (in GHz). 
We estimate the average magnetic field strength in the emitting region by assuming the equipartition condition, in which the total energy is minimized and the energy densities of relativistic particles and magnetic fields are approximately equal  \citep{govoni04}. Then the equipartition magnetic field strength is given by:

\begin{align}
    B_{\rm eq} = & \left[ \frac{24\pi}{7} \xi(\alpha, \nu_1, \nu_2) (1+k)^{4/7} (\nu_{0[\rm MHz]})^{4\alpha/7} \right. \nonumber \\
    & \left.  (1+z)^{(12+4\alpha)/7} (I_{0[\frac{\rm mJy}{\rm arcsec^2}]})^{4/7} (d_{\rm [kpc]})^{-4/7} \right]^{1/2}
\end{align}

\noindent where $\alpha$ is the spectral index, $\xi(\alpha, \nu_1, \nu_2)$ is the equipartition parameter tabulated in Table~1 of \citet{govoni04} for the frequency range from 10~MHz to 100~GHz, $k$ is the ratio of energy in relativistic protons to that in electrons, $z$ is the redshift, $I_{0}$ is the total flux at frequency $\nu_0$ (in mJy), $\Omega$ is the solid angle of the emitting region (in sr), and $d$ is the source depth (in kpc). 
We adopt $\xi(\alpha, \nu_1, \nu_2)=1.23\times10^{-12}$, which is given for $\alpha\sim0.7$ between the turnover and break frequencies \citep{govoni04}, and the total flux of $I_{0}$=7.297~Jy at $\nu_0$=4.885~GHz\footnote{$B_{\rm eq}$ varies by less than $\sim$6\% when using other values of $I_0$ at frequencies between 1--40~GHz in Table~9 in \citet{perley13_flux}.} \citep{perley13_flux}. Given that the total flux of 3C~286 at this frequency is dominated by emission within 50~mas from the core \citep[e.g.,][]{zhang94, jiang96, cotton97, an17}, we adopt a source size of 50~mas. This corresponds to a depth of $d=3.6\times10^{-1}$~kpc, assuming a spherical geometry for simplicity. 
Assuming an energy ratio between electrons and protons of $k=1$, as commonly adopted in the literature, these parameters yield an equipartition magnetic field strength of $B_{\rm eq}$$\sim$4.4~mG\footnote{$k$ can vary between 1 and 1836, depending on the locations or conditions of the emitting regions \citep[e.g.,][]{kino12, kawakatu16}. However, adopting $k=1836$ changes $B_{\rm eq}$ by only several factors ($B_{\rm eq}=31$~mG), indicating that $B_{\rm eq}$ is not highly sensitive to the choice of $k$.}. This value appears to align with those observed in other sources. For example, the magnetic field strengths of 4.4~mG are smaller than the magnetic field strength of $10^{1-4}$~G typically found in the vicinity of the supermassive black holes \citep[e.g.,][]{kino15, ehtc21, paraschos21, paraschos24, ro23, kino24, lisakov25}, but larger than or comparable to those found in radio lobes \citep[e.g.,][]{hardcastle02, murgia03, isobe15}. These results suggest that our estimated equipartition magnetic field strength is reasonable, as it reflects an average value derived from the emission across the entire jet of 3C~286. 

Substituting the magnetic field strength of 4.4~mG into Equation~\ref{eq:tau_syn}, we derive the age of 3C~286 to be $\tau_{\rm syn}$=450~years. Given that the source size of 50~mas corresponds to 362~pc, this age implies an average apparent speed $\beta_{\rm app}=2.6$. 
However, these apparent speeds are significantly faster than the apparent speed $\beta_{\rm app}=0.6\pm0.5$ observed in the inner jet at 5~mas from the core \citep{an17}. Moreover, this is also faster than the typical speeds observed in CSS sources, which are generally on the order of $\beta_{\rm app}=0.1$ \citep[e.g.,][]{giroletti09}. This discrepancy suggests that the spectrum may not represent the age of the entire jet, but instead may partially represent the age of electrons that were accelerated in situ as the jet propagates through the ambient medium. This, in turn, implies that the overall spectrum may not be fully explained by a one-zone synchrotron model and may instead require a more complex, multi-zone interpretation. For example, the break at 67~GHz may be predominantly caused by the inner jet, which has a size of $\sim$5~mas, rather than by the entire jet structure. If so, the discrepancy can be reconciled by introducing an additional break frequency at $\sim$3.6~GHz (Figure~\ref{fig:spectral}). Specifically, if the 67~GHz break traces synchrotron aging in the inner jet while the 3.6~GHz break corresponds to the outer jet at 50~mas, the inferred synchrotron age of the outer jet becomes 1950~years. This corresponds to an average apparent speed of $\beta_{\rm app}=0.6$, which is more consistent with the speed observed in the inner jet of 3C~286 and in other CSS sources. Future VLBI monitoring could help test this scenario by spatially resolving the spectral evolution along the jet.

\subsection{EVPA rotation as a function of frequency}

As shown in Section~\ref{sec:result_pol}, our monitoring of 3C~286 revealed that its EVPA gradually rotates as a function of frequency (Figure~\ref{fig:bothpol}). We suggest that this frequency dependence can be explained by the opacity effect. According to the Blandford \& K{\"o}nigl (BK) conical jet model, the core is defined as the region where the jet plasma becomes optically thin, i.e., where the optical depth reaches $\tau\sim1$ \citep{bk79}. In this model, the observed core position $r_{\nu}$ shifts toward the physical jet base at higher frequencies $\nu$, following $r_{\rm \nu}\propto \nu^{-k_{\rm r}}$; this phenomenon is referred to as core-shift \citep{lobanov98}. Core-shift has been widely observed in various types of AGN jets including radio galaxies \citep[e.g.,][]{hada11, haga15, paraschos21, park21}, blazars \citep[e.g.,][]{osullivan09_cs, okino22, yi24}, and narrow-line Seyfert~1 galaxies \citep{hada18}.

Core-shift is also observed in 3C~286. Previous VLBI observations consistently identified two bright knots within $\sim$5~mas from the core: one is C1, which is considered as the VLBI core, and the other one is C2, the closest downstream component located at 5~mas from C1 \citep[e.g.,][]{zhang94, jiang96, cotton97, nagai16, an17}. Notably, the separation between C1 and C2 becomes larger at higher frequencies between 2--15~GHz, suggesting that the position of C1 shifts closer to the black hole as frequency increases \citep{nagai16}. If the direction of the magnetic field gradually changes with optical depth, it could explain the observed EVPA rotation with frequency \citep{agudo12, nagai16}. 

\begin{figure}[t!]
\centering 
\includegraphics[scale=0.6]{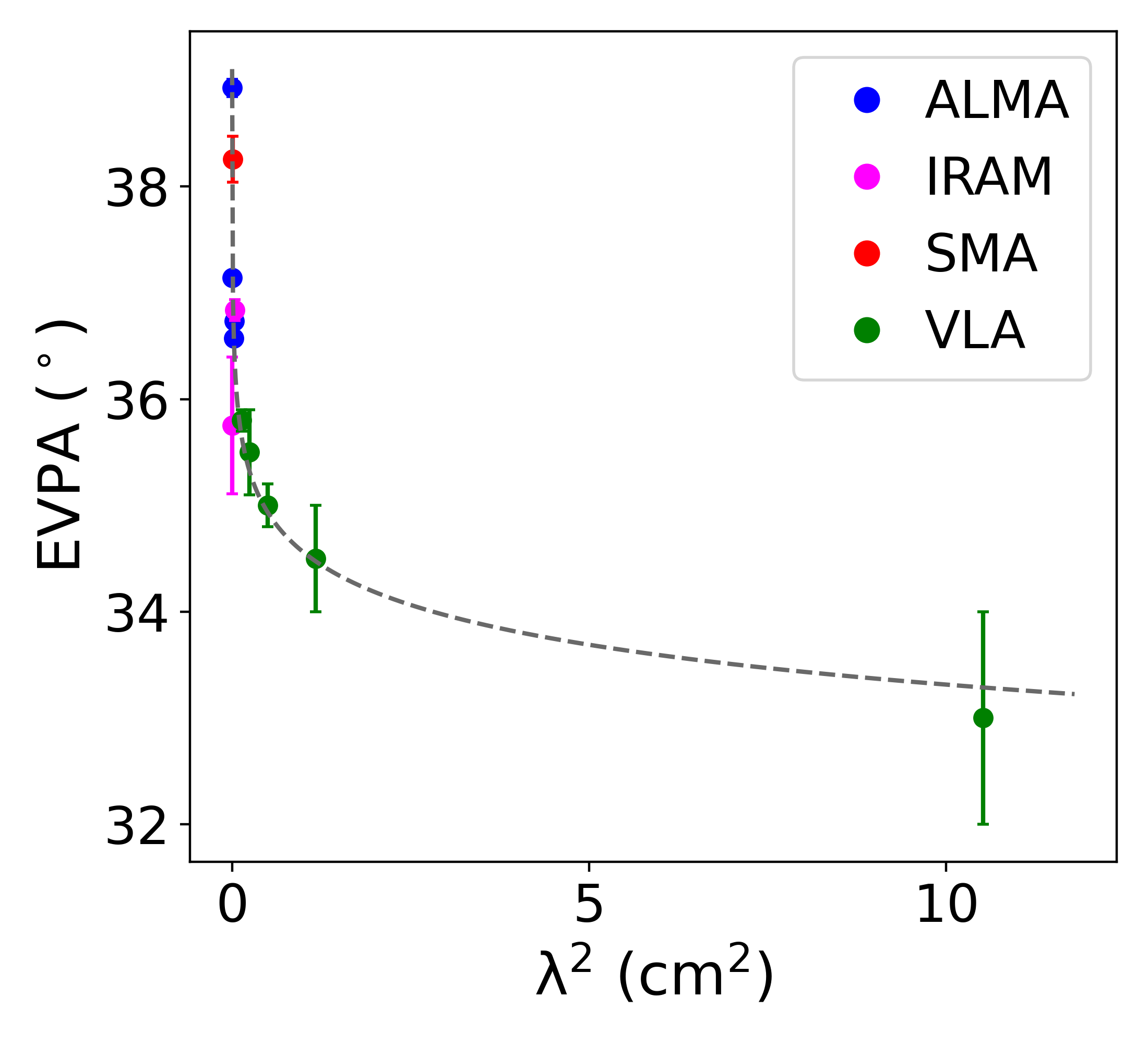}
\caption{The blue, magenta, red, and green dots represent the EVPA measurements from ALMA, IRAM, SMA, and VLA, plotted as a function of wavelength squared, respectively. The gray dashed line represents the best-fit to the EVPA values. 
\label{fig:evpa_fit}}
\end{figure}

In addition to this explanation, we should consider Faraday rotation, which causes the EVPA to rotate as a function of frequency when it passes through magnetized plasma. The amount of rotation is given by $\Delta\chi$ = RM$\lambda^2$, where $\lambda$ is the observing wavelength and RM is Faraday rotation measure. The RM is given by RM\,$=8.1\times10^5\ \int_{\rm LoS} n_{e} B\ dl$, where $n_e$ is the electron density (in cm$^{-3}$), $B$ is the magnetic field strength (in Gauss), and $l$ is the path length (in parsec) along the line of sight \citep{rybicki79, trippe14}. If the core-shift effect is observed, the RM at the core is expected to increase as a function of frequency, following $|{\rm RM}| \propto \nu^{a}$ with $n_{\rm e} \propto d^{-a}$, where $\nu$ is the observing frequency, $a$ is a power-law index in the electron density distribution, and $d$ is the distance from the jet base \citep[e.g.,][]{jorstad07}. This is because we would see a region closer to the jet base at higher frequencies due to the core-shift effect, where one may expect higher particle densities and stronger magnetic field strength. In case of the BK conical jet expanding conically, we may expect (i) a power-law electron density distribution as a function of distance from the core $n_{e}\propto d^{-a}$ with $a=2$, (ii) a path length proportional to the distance from the core $dl\propto d$, (iii) dominance of the toroidal magnetic field $B\propto d^{-1}$, and (iv) energy equipartition $d_{\rm core}\propto \nu^{-1}$, leading to $|{\rm RM}| \propto \nu^{a}$ with $a=2$ \citep[e.g.,][]{lobanov98, jorstad07, park18, hovatta19}. 

We present the EVPA values\footnote{We did not include $\chi=29.2^\circ$ at 1.4~GHz obtained with MeerKAT in the RM calculation, due to its frequency-dependent uncertainties of up to 3$^\circ$ (see Figure~2 in \citet{taylor24}). Including this measurement could introduce a systematic deviation in the fit and lead to an inaccurate RM estimate, particularly given that it lies orders of magnitude farther in wavelengths squared compared to the other data points.} of 3C~286 measured with VLA, ALMA, IRAM, and SMA as a function of the square of the wavelength in Figure~\ref{fig:evpa_fit}. This allows us to examine how the RM varies with frequency, as the slope between two data points in the EVPA-$\lambda^2$ plot is proportional to the RM. The EVPA of 3C~286 barely rotates as a function of the square of the wavelength at centimeter wavelengths, especially at frequencies below 10~GHz \citep{perley13_pol}. This suggests that the RM of 3C~286 is nearly negligible at these frequencies, as indicated by the relation $\rm \Delta\chi \propto \lambda^2 RM$. However, the EVPA gradually rotates from 33$^\circ$ to 39$^\circ$ as frequency increases from 5 to 343.4~GHz. In particular, the EVPA tends to rotate more rapidly at shorter wavelengths, indicating more rapid increase in RM. This behavior is consistent with what is expected from core-shift. 

\begin{figure}[t!]
\centering 
\includegraphics[scale=0.6]{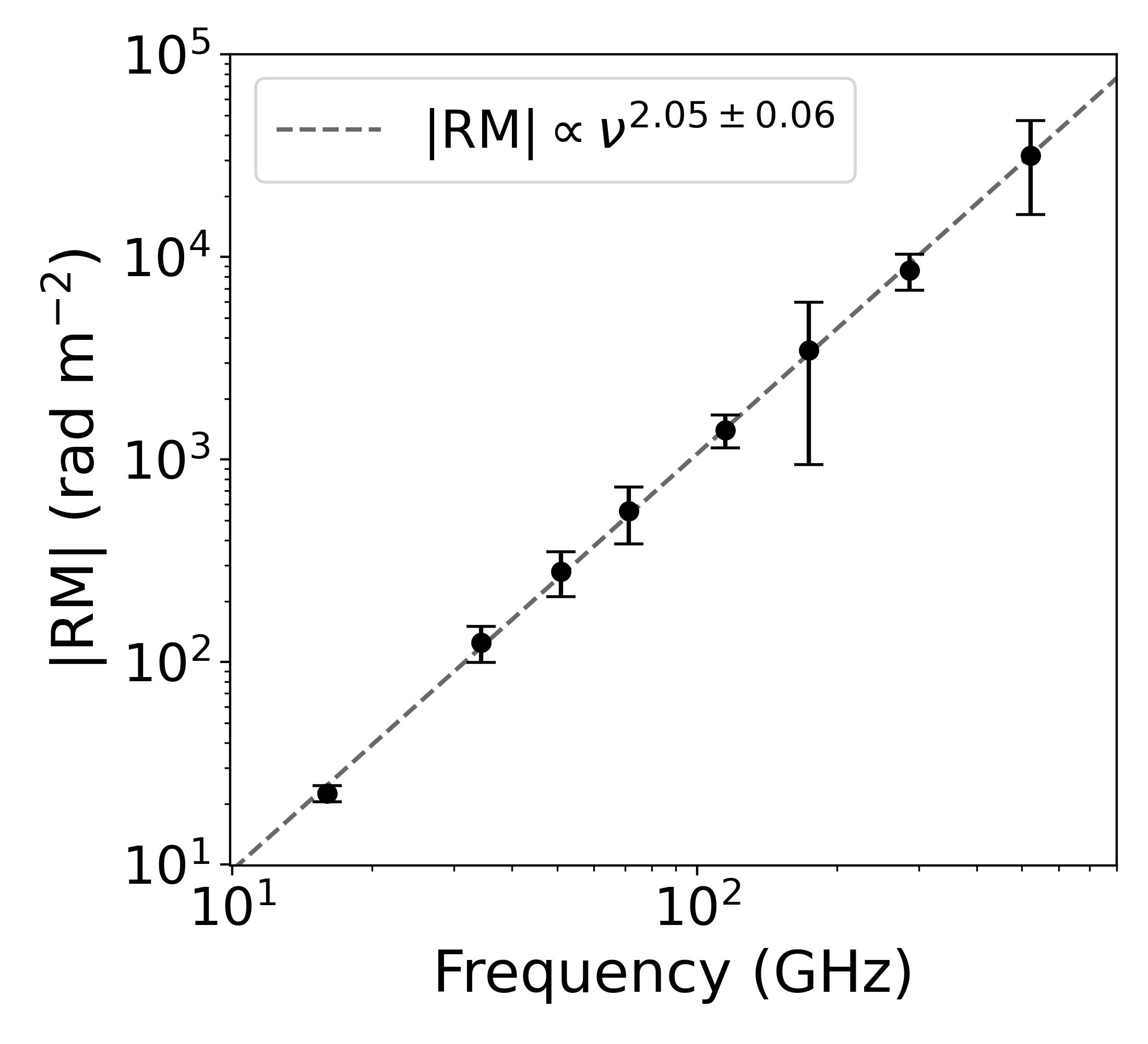}
\caption{RM values as a function of geometrical mean frequency for 3C~286 in its rest frame. Each RM value was calculated using the best-fit EVPA values at adjacent VLA and ALMA frequencies, plotted at the geometrical mean frequency of the pairs. The gray dashed line represents the best-fit to the RM values. 
\label{fig:rmfit}}
\end{figure}

\begin{table*}[t]
    \centering
    \setlength{\tabcolsep}{8pt}
    \caption{The RM of 3C~286 calculated from the best-fit EVPA at representative frequencies.}
    \begin{tabular}{c c c c c c}
    \toprule
     $\nu_1$ & $\nu_2$ & $\nu_{\rm mean,obs}$ & $\rm \left|RM_{\rm obs}\right|$ & $\nu_{\rm mean,rest}$ & $\rm \left|RM_{\rm rest}\right|$ \\ 
    $[$GHz$]$ & [GHz] & [GHz] & [rad m$^{-2}$] & [GHz] & [rad m$^{-2}$] \\
    \midrule
    \raggedright
    5 & 15 & 8.7 & $(6.5\pm0.7)\times10^0$ & 16.0 & $(2.3\pm0.2)\times10^1$ \\ 
    15 & 23 & 18.6 & $(3.6\pm0.8)\times10^1$ & 34.3 & $(1.2\pm0.3)\times10^2$ \\
    23 & 33 & 27.5 & $(8.0\pm1.9)\times10^1$ & 50.9 & $(2.8\pm0.7)\times10^2$ \\ 
    33 & 45 & 38.5 & $(1.6\pm0.5)\times10^2$ & 71.3 & $(5.6\pm1.7)\times10^2$ \\
    45 & 86 & 62.2 & $(4.0\pm0.6)\times10^2$ & 115.0 & $(1.4\pm0.3)\times10^3$ \\
    86 & 103 & 94.1 & $(9.9\pm6.4)\times10^2$ & 174.0 & $(3.5\pm2.6)\times10^3$ \\ 
    103 & 233 & 154.9 & $(2.5\pm0.4)\times10^3$ & 286.4 & $(8.6\pm1.7)\times10^4$ \\ 
    233 & 343 & 282.7 & $(9.1\pm3.8)\times10^3$ & 522.7 & $(3.2\pm1.5)\times10^4$ \\ 
    \bottomrule
    \end{tabular}
    \raggedright
    \tablecomments{The RM values obtained between adjacent pairs of frequencies. $\nu_{\rm mean,obs}$ and $\nu_{\rm mean,rest}$ represent the geometrical mean frequency of the adjacent frequency pairs in the observing and rest frames, respectively. $\rm \left|RM_{\rm obs}\right|$ and $\rm \left|RM_{\rm rest}\right|$ indicate the absolute values of the RM in the observing and rest frame frequencies, respectively.} 
    \label{tab:rm_values}
\end{table*}

To investigate the frequency dependency of the RM in more detail, we fit all the EVPA values with a single power-law function of the form

\begin{equation}
    \mathrm{log(\chi)} = k\ \mathrm{log(\lambda^2) + C}
\end{equation}

\noindent where $\chi$ is the EVPA, and $\lambda$ is the wavelength in centimeter (Figure~\ref{fig:evpa_fit}). The resulting coefficients from the weighted fit are $k=-0.016\pm0.002$ and $\rm C=1.547\pm0.003$. Using these coefficients, we sample the EVPA values along the best-fit curve at ten representative frequencies and calculate the best-fit RM values between adjacent pairs. Since these model EVPAs along the best-fit curve are derived from a fit to multiple data points, their statistical uncertainties are smaller than those of individual EVPA measurements by a factor of the square root of the number of EVPA measurements. We find that the absolute value of the RM between 5 and 343~GHz increases from $\rm 6.5\ rad\ m^{-2}$ to $\rm 9100\ rad\ m^{-2}$ in the observing frame. This corresponds to an increase from $\rm 23\ rad\ m^{-2}$ to $\rm 32000\ rad\ m^{-2}$ in the rest frame, where the rest frame RM is defined as 

\begin{equation}
    \rm \left|RM_{\rm rest}\right| = (1+z)^2 \left|RM_{\rm obs}\right| 
\end{equation}

\noindent with $\rm \left|RM_{\rm rest}\right|$ and $\rm \left|RM_{\rm obs}\right|$ denoting the rest frame and the observing frame RMs, respectively (Table~\ref{tab:rm_values}). These RM values are significantly larger than the previously reported values for 3C~286, which were believed to be close to zero, but falls within the typical range observed in CSS sources \citep[see][]{taylor92, luedke98, mantovani09, mantovani13}. 


In Figure~\ref{fig:rmfit}, we present the absolute values of RM as a function of frequency in the rest frame. The best-fit line represents $\mid$RM\textsubscript{core,$\nu$}$\mid\propto \nu^{a}$ with the power-law index $a=2.05\pm0.06$, which is consistent with the theoretically expected value of $a=2$ for the BK conical jet model. This supports the interpretation that the observed EVPA rotation as a function of wavelength squared is primarily due to the core-shift effect in a conical jet \citep{nagai16}.

Given that the observed EVPA values are not only well-described by a single power-law function over three orders of magnitude in wavelength squared, but are also consistent with a straightforward physical explanation (core-shift effect in a conical jet), we propose using the theoretical EVPA values from this model curve for absolute EVPA calibration between 5 and 343.4~GHz.

\subsection{Depolarization mechanism}

\begin{figure}[t!]
\centering 
\includegraphics[scale=0.6]{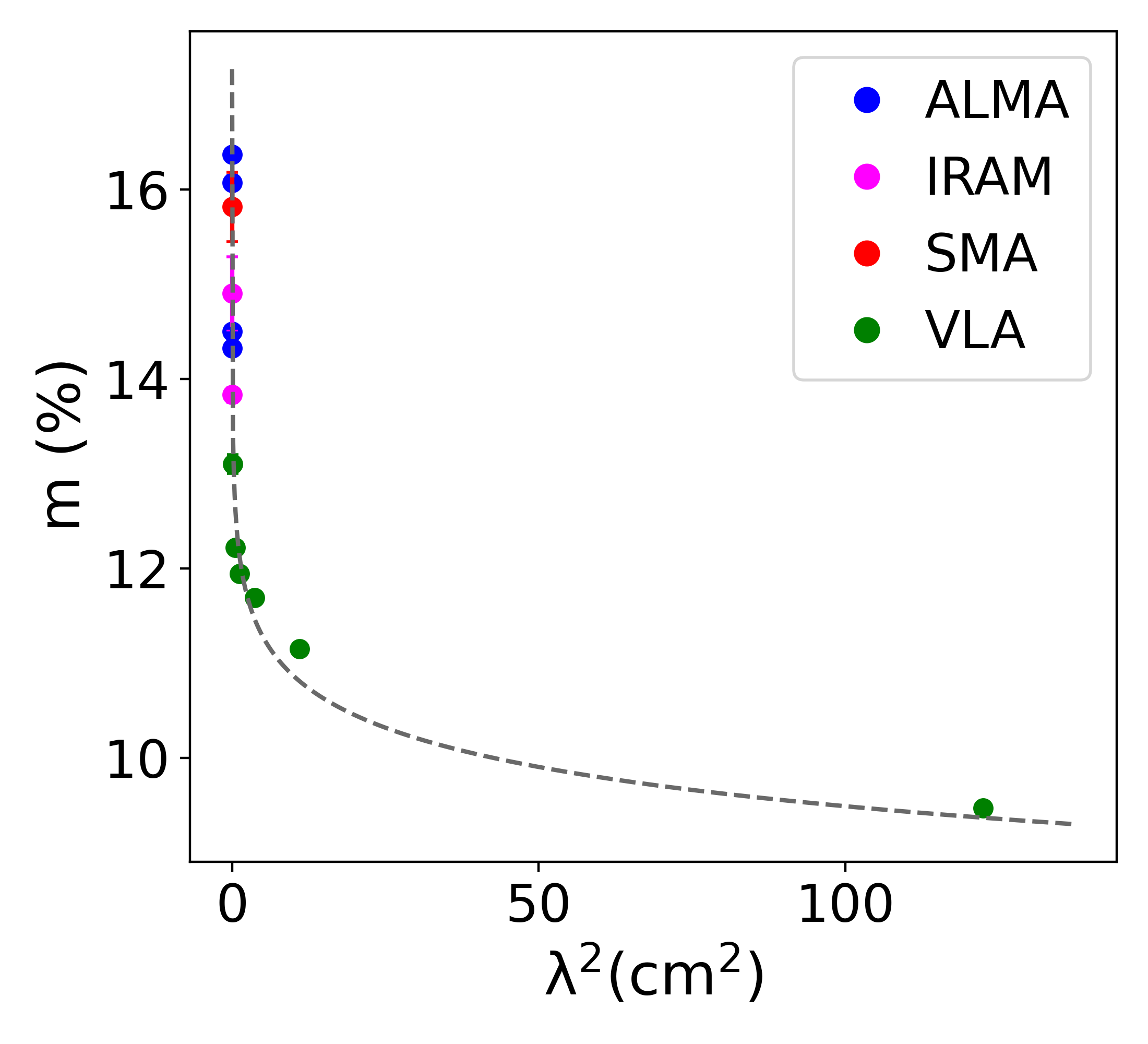}
\caption{The blue, magenta, red, and green dots represent the fractional polarization measured with ALMA, IRAM, SMA, and VLA, plotted as a function of wavelength squared. The gray dashed line represents the best-fit to the fractional polarization values.
\label{fig:m_fit}}
\end{figure}

In Figure~\ref{fig:m_fit}, we present the fractional polarization of 3C~286 as a function of wavelength squared from 1.5~GHz to 343.4~GHz. The fractional polarization increases as the wavelength decreases, with the rate of increase is more rapid at shorter wavelengths. This behavior closely mirrors the EVPA rotation, which also becomes more rapid at shorter wavelengths (Figure~\ref{fig:evpa_fit}). Such a trend can be naturally explained with the reduced effect of depolarization due to Faraday rotation, as we explain in the following. 

If the foreground plasma is not uniform within an instrumental resolution element, the EVPA passing through different plasma regions will undergo different magnitudes of Faraday rotation, resulting in depolarization when these different EVPAs are averaged within the instrumental beam \citep[e.g.,][]{burn66, sokoloff98, pasetto18}. Consequently, higher fractional polarization is expected at shorter wavelengths, where Faraday rotation is less effective. This is because EVPA rotates following $\rm \Delta\chi \propto \lambda^2 RM$. Although RM increases at shorter wavelengths, the magnitude of the increase in RM is smaller than the decrease in wavelength squared, resulting in a reduced depolarization effect. For example, as $\lambda^2$ decreases by a factor of 4706, from 5 to 343~GHz, the RM$_{\rm obs}$ increases by only $1400$, from 6.5 to 9100~rad~m$^{-2}$. This indicates that the decrease in wavelength squared is larger than the increase in RM, suggesting that Faraday depolarization is reduced at shorter wavelengths. These observation results align well with depolarization caused by Faraday rotation. 

Motivated by the EVPA distribution, we fit the fractional polarization values with a single power-law function of the form: 
\begin{equation}
    \mathrm{log}(m) = k'\ \mathrm{log(\lambda^2) + C'}
\end{equation}
\noindent where $m$ denotes the fractional polarization and $\lambda$ is the wavelength in centimeters. Notably, the fractional polarization, spanning four orders of magnitude in wavelength squared, is well-described by a single power-law function with coefficients $k'=-0.047\pm0.001$ and $\rm C'=1.108\pm0.001$ (Figure~\ref{fig:m_fit}). This closely resembles the EVPA distribution, which is also well fitted with a single power-law, suggesting that at shorter wavelengths, less efficient Faraday depolarization entails the increase of the fractional polarization.

We note that various depolarization models which relate fractional polarization to wavelength have been studied in both theoretical studies \citep[e.g.,][]{burn66, cioffi80, tribble91, sokoloff98, melrose98, yushkov24} and observational analyses \citep[e.g.,][]{mantovani09, farnes14, pasetto18, pasetto21, park19_rm} to explore the properties of the region where Faraday rotation occurs, such as whether the rotation is internal or external to the emitting region, the geometry of the region (e.g., slab or sphere), the number of plasma layers, and complex structure of the magnetic fields within the plasma. However, those models are primarily suited for optically thin cases, whereas the emitting regions in 3C~286 are partially optically thick, exhibiting the core-shift effect. In addition, the higher fractional polarization observed at shorter wavelengths may also contribute to more ordered magnetic fields in regions closer to the black hole \citep{nagai16}, further complicating the investigation of the Faraday rotating media. We expect that future theoretical studies based on our observations or polarimetric observations with VLBI at millimeter/submillimeter will provide a more detailed understanding of the origin of Faraday rotation in 3C~286 by resolving the optically thin and thick regions of the jet.

\section{Conclusions} \label{sec:Conclusions}
In this paper, we have reported the results from the monitoring observations of 3C~286 using three instruments: ALMA at 91.5, 103.5, 233.0, and 343.4~GHz, IRAM at 86 and 229~GHz, and SMA at 225~GHz. 

The total flux measured with IRAM is nearly constant throughout the monitoring period from 2006 to 2024, whereas the ALMA flux measurements show decreasing trends during its monitoring period from 2018 to 2024. This discrepancy can be partially related to a 4\% increase of the brightness of Uranus, which serves as the primary flux calibrator for ALMA. This increase may result from the gradual rotation of Uranus' north pole into our view, as it is brighter than the equatorial regions, thereby increasing the integrated brightness temperature of Uranus. However, the ESA4 model used in ALMA flux calibration does not take into account such time-dependent variations of brightness. Instead, ALMA also monitors other SSOs such as Neptune and Mars, and adopts flux scaling factors derived from them only when they agree with the scaling factor from Uranus within 5\%. As a consequence, the flux of other sources including 3C~286 may likely be scaled down within these limits, which could lead to an apparent gradual decrease in the flux of 3C~286 measured with ALMA. Therefore, we conclude that the flux of 3C~286 has remained nearly constant within the IRAM measurement uncertainties over the past $\sim$20 years.

The average total flux measured with IRAM at 86 and 229~GHz corresponds to a spectral index of $1.039\pm0.008$, which is steeper than the spectral index of $0.712\pm0.003$ observed between 3 and 50~GHz. This steeper spectrum at higher frequencies can be explained by synchrotron cooling - a process in which higher-energy electrons lose energy more rapidly through synchrotron radiation than lower-energy electrons. As a result, the spectrum exhibits a break at $67.2\pm1.2$~GHz. 

The age of the jet, derived from the equipartition magnetic field strength of 4.4~mG and the break frequency of 67.2~GHz, is 450~years under the \textit{continuous injection} model. This corresponds to an average apparent speed of $\beta_{\rm app}=2.6$, given the size of the source of 50~mas. However, this is significantly faster than the typical speeds observed in CSS sources, implying that the spectrum may not be explained by a one-zone synchrotron model.

We have also investigated the polarimetric properties of 3C~286. Both the EVPA and fractional polarization are stable from 86~GHz to 343.4~GHz over the past $\sim$20~years. This is in agreement with the stability at frequencies lower than 50~GHz confirmed with previous studies. Notably, the EVPA gradually rotates with frequency, from $33.0^\circ\pm1.0^\circ$ at 5~GHz to $38.9^\circ\pm0.1^\circ$ at 343.4~GHz. 
This gradual EVPA rotation as a function of frequency can be explained by the core-shift effect. For example, a gradual change in the magnetic field direction with optical depth, as suggested by previous studies, could account for the EVPA rotation. 

Alternatively, it can also be explained with Faraday rotation, without requiring magnetic fields that gradually change direction with optical depth. Specifically, the EVPA rotation is observed to be more rapid at shorter wavelengths, indicating that the RM increases at shorter wavelengths. We performed a power-law fitting to the EVPA values from 5 to 343.4~GHz and found that the RM increases from 23 to $\rm 32000~rad~m^{-2}$ in the rest frame. This is the first time such an increase in the RM has been observed for 3C~286. Moreover, the RM increases with frequency, following a power-law relation of RM$_{\rm core} \propto \nu^{a}$ with $a=2.05\pm0.06$. This result aligns with the power-law index of $a=2$ expected for the core-shift effect in the BK conical jet model, suggesting that the observed EVPA rotation in 3C~286 can be explained in this framework.

Given that the EVPA values are not only fitted well with a single power-law over three orders of magnitude in wavelength squared, but also aligns with a physical explanation based on the core-shift effect in a conical jet, we propose using the theoretical EVPA values provided by the model curve for absolute EVPA calibration between 5 and 343.4~GHz. 

Meanwhile, the fractional polarization gradually increases with frequency, from $8.6\%$ at 1.1~GHz to $16.4\%\pm0.1\%$ at 343.4~GHz. Notably, it follows a single power-law as a function of wavelength squared, similar to the trend found in the EVPA distribution. This implies that the higher fractional polarization at shorter wavelengths may result from reduced depolarization caused by Faraday rotation, as its effect diminishes at shorter wavelengths. In addition, more ordered magnetic fields in regions closer to the central black hole could also contribute to the higher fractional polarization at shorter wavelengths.

\begin{acknowledgments}
This research was supported by Basic Science Research Program through the National Research Foundation of Korea (NRF) funded by the Ministry of Education (RS-2024-00412117). We acknowledge financial support from the National Research Foundation of Korea (NRF) grant 2022R1F1A1075115. 
This work was supported by the National Research Foundation of Korea (NRF) grant funded by the Korea government (MSIT; RS-2024-00449206). This research has been supported by the POSCO Science Fellowship of POSCO TJ Park Foundation.
M.K. is supported by the JSPS KAKENHI Grant number JP22H00157 and JP21H01137.
K.A. acknowledges financial support from AS-CDA-110-M05, NSTC 113-2112-M-001-034, and NSTC 113-2124-M-001-008. 
CC and DA acknowledge support from the European Research Council (ERC) under the HorizonERCGrants2021 programme under grant agreement No. 101040021.
T.M. acknowledges the support by the National Science and Technology Council, Taiwan, under grant Nos. MOST 110-2112-M-001-068-MY3, 113-2112-M-001-028-, and NSTC114-2112-M-001-012- and the Academia Sinica, Taiwan, under a career development award under grant No. AS-CDA-111-M04.
This paper makes use of the following ALMA data: ADS/JAO.ALMA\#2011.0.00001.CAL. ALMA is a partnership of ESO (representing its member states), NSF (USA) and NINS (Japan), together with NRC (Canada), MOST and ASIAA (Taiwan), and KASI (Republic of Korea), in cooperation with the Republic of Chile. The Joint ALMA Observatory
is operated by ESO, AUI/NRAO and NAOJ. The IAA-CSIC group acknowledges financial support from the grant CEX2021-001131-S funded  by MCIN/AEI/10.13039/501100011033 to the Instituto de Astrof\'isica de Andaluc\'ia-CSIC". Acquisition and reduction of the POLAMI data was supported in part by MICIN through grants PID2019-107847RB-C44 abd PID2022-139117NB-C44. The POLAMI observations were carried out at the IRAM 30m Telescope. IRAM is supported by INSU/CNRS (France), MPG (Germany) and IGN (Spain). The Submillimeter Array (SMA) is a joint project between the Smithsonian Astrophysical Observatory and the Academia Sinica Institute of Astronomy and Astrophysics and is funded by the Smithsonian Institution and the Academia Sinica. Maunakea, the location of the SMA, is a culturally important site for the indigenous Hawaiian people; we are privileged to study the cosmos from its summit.

\end{acknowledgments}

\appendix

\section{The total flux, fractional polarization, and EVPA of 3C~286 measured with ALMA}\label{sec:appendix_flux}

In Table~\ref{tab:app_flux}, we present the average total flux, standard deviation, and average measurement uncertainty of the total flux of 3C~286 as measured with ALMA during Period~I, II, and III. Table~\ref{tab:app_alma_evpa} and \ref{tab:app_alma_fracpol} are the same as Table~\ref{tab:app_flux} but for EVPA and fractional polarization, respectively.

\begin{table}[t!]
    \setlength{\tabcolsep}{4pt}
    \renewcommand{\arraystretch}{}
    \caption{Total flux measurements with ALMA.}
    \begin{tabular}{c c c c c c c c c c c c c c c c}
    \toprule
     & \multicolumn{4}{c}{Period I} & & \multicolumn{4}{c}{Period II}& & \multicolumn{4}{c}{Period III} \\ 
    \cmidrule{2-5}
    \cmidrule{7-10}
    \cmidrule{12-15}
     $\nu$ & N & $S_\nu$ & $\sigma_{S_\nu}$ & $\langle \varepsilon_{S_\nu} \rangle$ & 
     & N & $S_\nu$ & $\sigma_{S_\nu}$ & $\langle \varepsilon_{S_\nu} \rangle$ & & N & $S_\nu$ & $\sigma_{S_\nu}$ & $\langle \varepsilon_{S_\nu} \rangle$ \\ 
    $[$GHz$]$ & (1) & (2) & (3) & (4) & & (1) & (2) & (3) & (4) & & (1) & (2) & (3) & (4) \\    
    \midrule
    91.5 & 53 & 0.812 & 0.036 & 0.034 & & 65 & 0.805 & 0.020 & 0.031 & & 62 & 0.788 & 0.025 & 0.032 \\  
    103.5 & 54 & 0.723 & 0.024 & 0.034 & & 62 & 0.719 & 0.018 & 0.029 & & 49 & 0.709 & 0.027 & 0.033 \\ 
    233.0 & 7 & 0.320 & 0.019 & 0.033 & & 12 & 0.311 & 0.014 & 0.027 & & 11 & 0.302 & 0.012 & 0.024 \\ 
    343.4 & 32 & 0.220 & 0.021 & 0.021 & & 36 & 0.205 & 0.016 & 0.021 & & 51 & 0.193 & 0.018 & 0.043 \\ 
    \bottomrule
    \end{tabular}
    \tablecomments{Total flux measurements in units of Jy. Period I, II, and III cover MJD 58200 to 59000, MJD 59200 to 59900, and MJD after 60100, respectively. The columns indicate (1) the number of data, (2) total flux in units of Jy, (3) standard deviation of total flux, and (4) average measurement uncertainty.} 
    \label{tab:app_flux}
\end{table}

\begin{table}[t]
    \centering
    \setlength{\tabcolsep}{4pt}
    \caption{EVPA measurements using ALMA.}
    \begin{tabular}{c c c c c c c c c c c c c c c c}
    \toprule
     & \multicolumn{4}{c}{Period I} & & \multicolumn{4}{c}{Period II} & & \multicolumn{4}{c}{Period III} \\ 
    \cmidrule{2-5} 
    \cmidrule{7-10} 
    \cmidrule{12-15} 
     $\nu$ & N & $\chi$ & $\sigma_{\chi}$\tnote{a} & $\langle \varepsilon_{\chi} \rangle$\tnote{b} & & 
     N & $\chi$ & $\sigma_{\chi}$\tnote{a} & $\langle \varepsilon_{\chi} \rangle$\tnote{b} & & 
     N & $\chi$ & $\sigma_{\chi}$\tnote{a} & $\langle \varepsilon_{\chi} \rangle$\tnote{b} \\ 
     $[$GHz$]$ & (1) & (2) & (3) & (4) & & 
     (1) & (2) & (3) & (4) & & 
     (1) & (2) & (3) & (4) \\
    \midrule
    91.5 & 72 & 36.87 & 1.12 & 1.23 & & 
    124 & 36.43 & 1.54 & 1.42 & &
    78 & 35.75 & 2.36 & 2.03 \\ 
    103.5 & 72 & 36.77 & 1.38 & 1.40 & & 
    124 & 36.90 & 1.47 & 1.59 & &
    78 & 34.87 & 2.37 & 2.44 \\ 
    233.0 & 9 & 36.89 & 2.00 & 2.02 & & 
    34 & 36.14 & 3.28 & 2.19 & &
    15 & 37.14 & 1.16 & 1.63 \\ 
    343.4 & 64 & 38.65 & 8.06 & 4.95 & & 
    111 & 38.13 & 5.26 & 3.91 & & 
    53 & 38.31 & 2.53 & 2.97 \\  
    \bottomrule
    \end{tabular}
    \raggedright
    \tablecomments{Same as Table~\ref{tab:app_flux} but for EVPA.} 
    \label{tab:app_alma_evpa}
\end{table}

\begin{table}[t]
    \centering
    \setlength{\tabcolsep}{4pt}
    \caption{Fractional polarization measurements using ALMA.}
    \begin{tabular}{c c c c c c c c c c c c c c c c}
    \toprule
     & \multicolumn{4}{c}{Period I} & & \multicolumn{4}{c}{Period II} & & \multicolumn{4}{c}{Period III} \\ 
    \cmidrule{2-5} 
    \cmidrule{7-10} 
    \cmidrule{12-15} 
     $\nu$ & N & $m$ & $\sigma_{m}$\tnote{c} & $\langle \varepsilon_{m} \rangle$\tnote{d} & & 
     N & $m$ & $\sigma_{m}$\tnote{c} & $\langle \varepsilon_{m} \rangle$\tnote{d} 
     & & 
     N & $m$ & $\sigma_{m}$\tnote{c} & $\langle \varepsilon_{m} \rangle$\tnote{d} \\ 
     $[$GHz$]$ & (1) & (2) & (3) & (4) & & 
     (1) & (2) & (3) & (4) & & 
     (1) & (2) & (3) & (4) \\
    \midrule
    91.5 & 72 & 14.24 & 0.60 & 0.66 & & 
    124 & 14.39 & 0.90 & 0.79 & &
    78 & 14.29 & 0.67 & 0.94 \\ 
    103.5 & 72 & 14.37 & 0.83 & 0.75 & & 
    124 & 14.70 & 1.41 & 0.90 & &
    78 & 14.25 & 0.97 & 1.10 \\ 
    233.0 & 9 & 15.55 & 0.93 & 1.03 & & 
    34 & 15.30 & 2.41 & 0.93 & &
    15 & 16.31 & 0.54 & 1.02 \\ 
    343.4 & 64 & 16.24 & 3.25 & 2.04 & & 
    111 & 16.01 & 2.58 & 1.78 & & 
    53 & 15.91 & 1.69 & 1.52 \\  
    \bottomrule
    \end{tabular}
    \raggedright
    \tablecomments{Same as Table~\ref{tab:app_flux} but for fractional polarization.} 
    \label{tab:app_alma_fracpol}
\end{table}

%

\vspace{5mm}

\bibliography{ref}{}
\bibliographystyle{aasjournal}

\end{document}